%% file: root_auto.tex
\documentclass[twocolumn]{autart}    

\usepackage{graphics} 

\usepackage{times} 
\usepackage{amsmath} 
\usepackage{amssymb}  
\usepackage[dvips]{psfrag}
\usepackage{psfrag}

\usepackage{url} 
\usepackage{mathtools}
\usepackage{tikz}
\usetikzlibrary{shapes,arrows}
\usepackage{verbatim}
\usepackage{standalone} 
\usetikzlibrary{positioning}

\newtheorem{theorem}{Theorem}

\newtheorem{proposition}{Proposition}[section]
\newtheorem{corollary}{Corollary}

\newtheorem{definition}{Definition}[section]
\newtheorem{example}{Example}

\DeclareMathOperator{\E}{\mathbb{E}}
\DeclareMathOperator{\Tr}{\text{Tr}}
\DeclareMathOperator{\T}{\rm T}
\DeclareMathOperator{\diag}{\mathrm{diag}}

\DeclareMathOperator{\HH}{\mathcal{H}}
\newcommand{\G}{{\mathcal{G}}}
\newcommand{\V}{{\mathcal{V}}}
\newcommand{\EE}{{\mathcal{E}}}
\DeclareMathOperator{\CT}{\text{H}}
\DeclareMathOperator{\N}{\mathnormal{n}}
\DeclareMathOperator{\M}{\mathnormal{d}}

\newcommand{\rhoo}{\rho_\text{ss}}

\newcommand{\BBTdiag}{b}
\newcommand*{\dbar}[1]{\overline{\overline{#1}}}

\newcommand{\HS}{\hspace{0.05cm}}
\newcommand{\NHS}{\hspace{-0.07cm}}

\begin{document}

\begin{frontmatter}

\title{ \bf
	Centrality in Time-Delay Consensus Networks with  \\ Structured Uncertainties 
}


\author[First]{Yaser Ghaedsharaf} \hspace{0.1cm}
\author[Second]{Milad Siami}  \hspace{0.1cm}
\author[First]{Christoforos Somarakis}  \hspace{0.1cm}
\author[First]{Nader Motee}

\address[First]{Department of Mechanical Engineering and Mechanics, Lehigh University, Bethlehem, PA 18015, USA\\ (e-mail: \tt \{ghaedsharaf, csomarak, motee\}@lehigh.edu).}
\address[Second]{The Institute for Data, Systems, and Society, Massachusetts Institute of
	Technology, Cambridge, MA 02139, USA.
	{ (e-mail:\tt siami@mit.edu)}}


\begin{abstract}
We investigate notions of network centrality in terms of the underlying coupling graph of the network, structure of exogenous uncertainties, and communication time-delay. Our focus is on time-delay linear consensus networks, where  uncertainty is modeled by structured additive noise on the dynamics of agents. The centrality measures  are defined using the $\HH_2$-norm of the network. We quantify the centrality measures  as functions of time-delay, the graph Laplacian, and the covariance matrix of the input noise. Several practically relevant uncertainty structures are considered, where we discuss two notions of centrality: one w.r.t intensity of the noise and the other one w.r.t coupling strength between the agents. Furthermore, explicit formulas for the centrality measures are obtained for all types of uncertainty structures. Lastly, we rank agents and communication links based on their centrality indices and highlight the role of time-delay and uncertainty structure in each scenario. Our counter intuitive grasp is that  some of centrality measures are highly volatile with respect to time-delay.
\end{abstract}

\end{frontmatter}

\section{Introduction}
Our objective is to study centrality of components of a network with respect to time-delay and coupling graph in presence of different sources of uncertainty. Measures of importance and influence in a network are called centrality measures and are exploited to provide a rank on the most important components of the network \cite{freeman1977set}.
Centrality is a well-studied subject in network analysis and graph theory and several measures are developed to address this importance in complex networks.
The degree centrality can be viewed as one of the simplest and most intuitive indices for centrality, which is defined as the number of connection that a node has to other nodes \cite{freeman1978centrality}. Betweenness centrality is based on shortest paths in a graph and its application vary from modeling traffic flows to telecommunication for ranking both nodes and links \cite{freeman1977set,girvan2002community}. Another popular class of indices for centrality is eigenvector centrality \cite{bonacich1972factoring,bonacich1987power}, which consists of PageRank \cite{page1999pagerank}. Other long-established centrality measures include Katz centrality \cite{katz1953new} and closeness centrality 
\cite{bavelas1950communication,sabidussi1966centrality}. Centrality indices for a noisy dynamical network was studied in \cite{siami2018centrality}. With time-delay being intrinsic to all real-world networks, this work studies effect of time-delay on centrality of nodes (agents) and edges (communication links) in a dynamical network. We borrow our notion of centrality from \cite{siami2018centrality}, where the authors define a performance measure and then quantify influence of each component of the network on the performance as their centrality index.

Measures of performance for linear consensus networks have been subject to extensive study  \cite{bamieh2008effect,Bamieh:2012,siami2016fundamental,young2010robustness,Zelazo:2011,moradian2018robustness}. Authors in \cite{bamieh2008effect} define a performance metric based on deviation from the average, while  \cite{siami2016fundamental} did a thorough study of this measure for first-order consensus networks. In \cite{yaser2016delay}, authors study $\HH_2$-based performance of the first-order linear network in presence of time-delay and show how interconnection topology can be designed to enhance performance via sparsification, adding new communication links, and feedback gain adjustment.  

We study a class of time-delay first-order consensus networks in the presence of noise input. Motivated by ideas from \cite{siami2018centrality}, we classify six types of uncertainty structures that appear in most real-world applications and we derive centrality indices as a function time-delay, the underlying graph, and structure of additive noise. The focus of this paper is on effect of time-delay on centrality  of individual  agents and communications links. We argue that increasing time-delay may shuffle centrality rankings. {In addition, we address critical role of connectivity in the presence of time-delay and compare it with the case that time-delay is absent.} This manuscript is an extension of \cite{ghaedsharaf2017eminence} that includes all the missing proofs of its conference version alongside new examples and materials in Sections \ref{sec:ranking} and \ref{sec:discussions} that are published for the first time. 
\section{Preliminaries and Definitions}

The  set of non-negative (positive) real numbers is indicated by $\mathbb{R}_{+}$ ($\mathbb{R}_{++}$). An undirected weighted graph $\mathcal{G}$ is denoted by the triple $\mathcal{G=(V,E,}w)$,  where $\mathcal{V}=\{v_1,v_2, \dots, v_{\N}\}$ is set of nodes (vertices) of the graph, $\mathcal{E}$ is set of links (edges) of the graph,  and $w: \EE \rightarrow \mathbb{R}_{++}$ is the weight function that maps each link to a positive scalar. We let $L$ to be the Laplacian of the graph, defined by 
$$L=\Delta-A,$$ where $\Delta$ is diagonal matrix of node degrees and $A$ is the adjacency matrix of the graph. Alternatively, we can write $L = E W E^{\T}$, where $E \in \mathbb{R}^{\N \times \mid \EE\mid}$ is the signed vertex-edge incidence matrix of the graph defined by
$$
[E]_{ie} = \begin{cases} +1 &\mbox{if } i \mbox{ is head of }e\\ 
-1 &\mbox{if } i \mbox{ is tail of }e\\ 
0 &\mbox{otherwise,}  \end{cases} 
$$
and $W \in \mathbb{R^{\mid \EE\mid \times \mid \EE\mid}}$ is the diagonal matrix of weights.\\
The $n \times 1$ vector of all zeros and ones  are denoted by $0_{\N}$ and $\mathbf{1}_{\N}$, respectively, while $J_{\N}=\mathbf{1}_{\N}\mathbf{1}_{\N}^{\T}$ is the $n \times n$ matrix of all ones. Conjugate transpose of matrix $G$ is denoted by $G^{H}$. Furthermore, the $n \times n$ centering matrix is denoted by $M_{\N}=I_{\N}-\frac{1}{\N}J_{\N}$. For an undirected connected graph with $\N$ nodes, Laplacian eigenvalues are real and  shown in an order sequence as $0=\lambda_1\leq\lambda_2
\leq\dots\leq \lambda_{\N}$.
We indicate Moore-Penrose pseudo-inverse of a Laplacian matrix $L$ by $L ^{\dagger} = [l_{ij}^{\dagger}]$ that can be utilized to define the effective resistance between nodes $i$ and $j$ using the following formula 
\[r_e(L) = l_{ii}^{\dagger}+l_{jj}^{\dagger}-2 l_{ij}^{\dagger}\]
for every given link $e=\{i,j\}$. For $X \in \mathbb{R}^{\N \times \N}$, the matrix-valued functions  $\cos (X)$ and $\sin(X)$ are defined as 

\begin{align*}
\cos(X) = \sum_{k=0}^{\infty}\frac{(-1)^k X^{2k}}{(2k)!},~~ \sin(X) = \sum_{k=0}^{\infty}\frac{(-1)^k X^{2k+1}}{(2k+1)!}.
\end{align*}

\section{Noisy Consensus Networks with Time-Delay}
In this paper, we consider a class of linear consensus networks whose dynamics are defined over graphs $\mathcal G = (\mathcal V, \mathcal E, w)$, where each node corresponds to a subsystem with a scalar state variable.
In study of consensus networks with delay, if a node has a delay in accessing or computing its state or has a delay in response, we add the self-delay to the model, i.e.,
\begin{align}\label{agentUpdate}
\dot{x}_i=\sum_{j \neq i} {a_{ij}\big({x_j(t-\tau)-x_i(t-\tau)}\big)},
\end{align}
where $a_{ij}$ is the $ij^\text{th}$ component of the adjacency matrix of the coupling graph. 

Therefore, the network that we study is the following single-delay consensus network with $\N$ nodes and underlying graph Laplacian $L$:

\begin{align}
\begin{aligned}
\label{eq:system}
\dot{x}(t)~&= -L\,x(t-\tau)+B\,\xi(t),\\
y(t)~&=~ M_{\N}\,x(t),
\end{aligned}
\end{align}
%
with $x(t)= 0$ for all $t \in [-\tau, 0)$ and $x(0)=x^0$, where  $x^0 = [x^0_1,  \ldots,  x^0_n]^{\rm T}$ is the initial condition, $x = [x_1,  \ldots,  x_n]^{\rm T}$ is the state, $y = [y_1,  \ldots,  y_n]^{\rm T}$ is the output, and $\xi = [\xi_1,  \ldots,  \xi_{\M}]^{\rm T}$ is the effect of an uncertain environment on agents or links. It is assumed that $\xi(t)$ is a vector of independent Gaussian white noise process with zero mean. The impact of an uncertain environment on each agent's dynamics is modeled by the exogenous noise input $\xi_i(t)$.  Furthermore, we assume that every agent experiences a time-delay in accessing, computing, or sharing its own state information with itself and other neighboring agents. It is assumed that the time-delay for all agents are identical and equal to a nonnegative constant $\tau$. The coupling graph of the consensus  network \eqref{eq:system} is a graph $\G=(\V,\mathcal E, w)$ with node set $\V=\{v_1,v_2,\ldots,v_{\N}\}$, edge set $\EE=\Big\{ \{i,j\}~\big|~\forall~i,j \in \V,~l_{ij} \neq 0\Big\}$ and weight function $w(\{i,j\})=-l_{ij}$ for all $e = \{i,j\} \in \EE$. The Laplacian matrix of graph $\G$ is equal to $L=[l_{ij}]$. 
\vspace{0.2cm}

For a consensus network, average consensus occurs if all agents converge to equal value in $\mathbb{R}$. For network \eqref{eq:system} it is known \cite{Olfati:2004} that this condition is equivalent to connectivity of the coupling graph and satisfying the following inequality  $$\tau \lambda_{\N} < \frac{\pi}{2}.$$

With goal of reaching an agreement among agents in a consensus network,  variance of  state of agents can be a measure of their performance. In other words, for a consensus network with underlying graph $L$ and in presence of time-delay $\tau$, can be measured by
\begin{align*}
\rhoo(L;\tau)\NHS \coloneqq\NHS \lim_{t\to \infty} \E \Big[\big(x(t) - \frac{1}{\N}J_{\N}\,x(t)\big)^{\T}\NHS \big(x(t)\NHS -\NHS \frac{1}{\N}J_{\N}\,x(t)\big)\NHS\Big],
\end{align*}
which can be equivalently be written as 
\begin{align*}
\rhoo(L;\tau) = \lim_{t\to \infty} \E \big[y(t)^{\T} y(t)\big].
\end{align*}

We utilize the notion of centrality introduced in \cite{siami2018centrality} in which authors study a consensus network in absence of time-delay and apply it to a time-delay first-order consensus network.
\begin{definition}
	For  network \eqref{eq:system}, let ${\xi_i(t) \sim \mathcal{N}(0,\sigma_i^2)}$ be the noise that is affecting agent $i$ for all $i \in \V $. Then, the centrality of agent $i$ is defined by
	\begin{align}\label{def1}
	\eta_i \coloneqq \frac{\partial \rhoo}{\partial \sigma_i^2}.
	\end{align}
\end{definition}
\vspace{0.2cm}
Here, $\eta_i$ measures the rate of change in the performance with respect to variance of the affecting noise on the agent $i$. In other words, it captures the effect of the disturbances associated with agent $i$ on the performance.

\vspace{0.2cm}
\begin{definition}
	For  network \eqref{eq:system}, let ${\xi_e(t) \sim \mathcal{N}(0,\sigma_e^2)}$ be the noise that is affecting the coupling $e$ for all $e \in \EE $. Then, the centrality of link $i$ is defined by
	\begin{align}\label{def2}
	\nu_e \coloneqq \frac{\partial \rhoo}{\partial \sigma_e^2}.
	\end{align}
\end{definition}
\vspace{0.2cm}
Thus, $\nu_e$ is the rate of change of the performance measure with respect to variance of the disturbance on the link $e$. Therefore, it represents outcome of the noise $e$ on the performance.
\vspace{0.2cm}
\begin{definition}
	For the consensus network \eqref{eq:system} and fixed time-delay $\tau \geq 0$ with a given structure for the input matrix $B$ and {identity} covariance  for the process $\xi(t)$, the {sensitivity coefficient} of the interconnection between nodes $i$ and $j$ is defined by
	\begin{align}\label{def3}
	\kappa_{e} \coloneqq \frac{\partial \rhoo}{\partial w(e)}.
	\end{align}
	\vspace{0.2cm}\\
	In other words, $\kappa_{e}$ is equal to derivative of the performance with respect to change in the weight of the interconnection. This quantity shows how much the performance will improve with a slight increase in weight of the interconnection.
\end{definition}
\begin{theorem}\label{th:Cent12_tau}
	Centrality indices $\eta_i$ and $\nu_e$ are increasing with respect to time-delay.
\end{theorem}
\begin{pf}
	Proof is straightforward by showing that the derivative of the centrality indices with respect to time-delay is positive in the stability region.
\end{pf}
\begin{theorem}\label{th:Cent12}
	For the consensus network \eqref{eq:system}, performance of the network can be written as a function of the Laplacian matrix $L$, uncertainty matrix $B$ and time-delay parameter $\tau$. In addition, we have the following identity
	\begin{align*}
	\rhoo(L,\tau) = \sum_{i \in \V} \eta_i\sigma_i^2
	\end{align*}
	with 
	\begin{align*}
	\eta_i = \frac{1}{2}\big[B^{T}L^{\dagger}\cos(\tau L)\big(M_n - \sin(\tau L)\big)^{\dagger}B\big]_{ii}.
	\end{align*}
	Moreover, we can obtain 
	\begin{align*}
	\rhoo(L,\tau) = \sum_{e \in \EE} \nu_e\sigma_e^2
	\end{align*}
	with 
	\begin{align*}
	{\nu_e} = \frac{1}{2}\big[B^{T}L^{\dagger}\cos(\tau L)\big(M_n - \sin(\tau L)\big)^{\dagger}B\big]_{ee}.
	\end{align*}
\end{theorem}
\begin{pf}
	We utilize the idea in \cite{siami2018centrality} for proof of this theorem. Let $\xi \in \mathbb{R}^{\M}$ in \eqref{eq:system}, then we define $\hat{\xi}_i \coloneqq {\xi}_i/\sigma_i$ for all $i\in \{1,\dots , \M \}$. From definition of $\xi_i$, we have
	\begin{align*}
	B \xi = \hat{B} \hat{\xi},
	\end{align*}
	where $\hat{B} = B \diag\big([\sigma_1,\dots,\sigma_{\M}]^{\T}\big)$. Consequently, we can write dynamics of the network \eqref{eq:system} as 
	\begin{align*}
	\dot{x}(t)~&= -L\,x(t-\tau)+\hat{B}\,\hat{\xi}(t),
	\end{align*}
	where from definition of $\hat{\xi}$ we note that $\hat{\xi}$ is a vector of unit variance and identically distributed Gaussian processes.
	
	In order to find the performance of the network (\ref{eq:system}), we utilize frequency domain definition of $\HH_2$-norm of the network \cite{Doyle89}, i.e.,
	\begin{align}\label{H2normCalc}
	\rhoo(L;\tau) &=& \frac{1}{2\pi}\Tr\Big[\int_{-\infty}^{+\infty}{G^{\CT}(j\omega)G(j\omega) \: d\omega}\Big]
	\end{align}
	with transfer matrix   
	\begin{align}\label{eq:Gs}
	G(s)~=~ M_{\N} \Big(sI_{\N}+e^{-\tau s}L\Big)^{-1}\hat{B}.
	\end{align}
	Although $G(s)$ is not exponentially stable, its single marginally stable mode is not observable in the output which consequently results in a bounded $\HH_2$-norm for the network.
	We consider spectral decomposition of Laplacian matrix $L$, which is,
	\begin{align*}
	L~=~Q \Lambda Q^{\T},
	\end{align*}
	where $Q=[q_1,q_2, \dots , q_{\N}] \in \mathbb{R}^{n\times n}$ is the orthonormal matrix of eigenvectors and $\Lambda=\diag(\lambda_1,\ldots,\lambda_{\N})$ is the diagonal matrix of eigenvalues. We recall that $\lambda_1=0$ for the reason that the graph is undirected and it has no self-loops.
	Therefore,
	\begin{align}
	M_{\N}~=&~I_{\N}-Q \diag([1,0, \dots,0]^{\T}) Q^{\T}\nonumber\\\label{eq:Mn}
	~=&~Q \diag([0,1, \dots,1]^{\T}) Q^{\T},
	\end{align}
	and
	\begin{align}
	L~=~Q \diag([0,\lambda_2, \dots, \lambda_{\N}]^{\T}) Q^{\T}. \label{eigen-decom}
	\end{align}
	Also, substituting \eqref{eq:Mn} and \eqref{eigen-decom} into (\ref{eq:Gs}), we obtain
	\begin{align*}
		G(s)~=~C Q \diag\NHS\Big(\big[0,\frac{1}{s+\lambda_2 e^{-\tau s}},\dots,\frac{1}{s+\lambda_{\N} e^{-\tau s}}\big]^{\T}\NHS\Big)Q^{\T}\NHS.
	\end{align*}
	Hence, we have
	\begin{align}\label{eq:GHG}
	&\Tr\big[G^{\CT}(j \omega)G(j \omega)\big]\nonumber\\
	=&\Tr\Bigg[ \hat{B}\hat{B}^{\T} Q \diag\Big(\Big[0,\frac{1}{\lambda_2 e^{j \tau  \omega}-j \omega},\dots,\frac{1}{\lambda_{\N} e^{j \tau \omega}-j \omega }\Big]^{\T}\Big)\nonumber\\
	\:\:&\diag\Big(\Big[0,\frac{d \omega}{j \omega+\lambda_2 e^{-j \tau  \omega}},\dots,\frac{1}{j \omega+\lambda_{\N} e^{-j \tau \omega}}\Big]^{\T}\Big)
	Q^{\T}\Bigg]
	\end{align}
	and by substituting (\ref{eq:GHG}) in \eqref{H2normCalc}, we obtain
	\begin{align*}
	\hspace{-0.15cm}\rhoo(L;\tau)
	= \frac{1}{2\pi} \sum_{i=2}^{\N}{\int_{-\infty}^{+\infty}\hspace{-0.3cm}\frac{\BBTdiag_{i} \HS d\omega}{\big(j \omega+\lambda_i e^{-j \tau \omega}\big)\big(\lambda_i e^{j \tau \omega}-j \omega\big)}}.
	\end{align*}
	where $\BBTdiag_{i}$ is the $i$'th diagonal element of the matrix  $Q^{\T}\hat{B}\hat{B}^{\T} Q$. 
	Simplifying the integral above
	we obtain 
	\begin{equation}
	\rhoo(L;\tau)=\sum_{i=2}^{\N} \frac{\BBTdiag_i}{2\lambda_i}~ \frac{\cos(\lambda_i \tau)}{1-\sin(\lambda_i \tau)}.\label{perf-meas}
	\end{equation}
	
	Now, we can  rewrite equality \eqref{perf-meas} in the following compact matrix  operator form
	\begin{align}
	\rhoo(L;\tau) = \frac{1}{2}\Tr\Big[&\hat{B}\hat{B}^{\T} L^{\dagger} \cos(\tau L)\Big(M_{\N}-\sin(\tau L)\Big)^{\dagger}\Big]\nonumber\\
	=\frac{1}{2}\Tr\Big[&\diag([\sigma_1^2,\dots,\sigma_{\M}^2]^{\T}){B}^{\T}\nonumber\\
	&L^{\dagger} \cos(\tau L)\Big(M_{\N}-\sin(\tau L)\Big)^{\dagger}{B}\Big].\label{perfFormula}
	\end{align}
	From identity \eqref{perfFormula} it is clear that when ${\M} = n$, i.e., ${B \in \mathbb{R}^{\N \times \N}}$ we obtain  
	\begin{align*}
	\eta_i = \frac{1}{2}\big[B^{T}L^{\dagger}\cos(\tau L)\big(M_n - \sin(\tau L)\big)^{\dagger}B\big]_{ii},
	\end{align*}
	for all $i \in {\V}$.
	Consequently, it follows that
	\begin{align*}
	\rhoo(L,\tau) = \sum_{i \in \V} \eta_i\sigma_i^2.
	\end{align*}
	Correspondingly, when existing links are affected by noises, i.e., $B \in \mathbb{R}^{\N \times \mid \EE \mid}$, we have
	\begin{align*}
	\nu_e = \frac{1}{2}\big[B^{T}L^{\dagger}\cos(\tau L)\big(M_n - \sin(\tau L)\big)^{\dagger}B\big]_{ee},
	\end{align*}
	for all $e \in \mathcal{E}$. Thus, in this case the following identity holds
	\begin{align*}
	\rhoo(L,\tau) = \sum_{e \in \EE} \nu_e\sigma_e^2.
	\end{align*}
\end{pf}

\section{Agent Associated Disturbances}
In this section, we consider four structures of disturbance for the network \eqref{eq:system} that affects agents and then find their centrality index. Structures of the noises arise from source of the noises that are affecting the system. In a consensus network within a noisy environment, 
uncertainties will appear in dynamics of the agent. Each individual agent, updates its state by sensing its own state, transmitting  its status to its neighbors, and receiving status of its neighbors. Since effect of uncertainties in each step of update affects the network in its sort of way, different types of uncertainty are modeled using different structures for the input matrix $B$.

\subsection{Dynamics Noise}

\begin{figure}
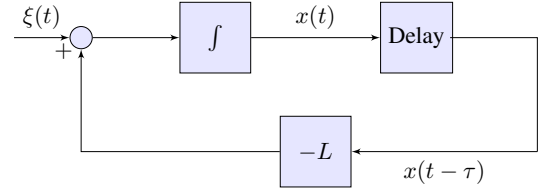

	\centering
	\includestandalone[width=0.42\textwidth]{dynamicNoise}
	\caption{Block diagram of consensus network \eqref{eq:system} in presence of dynamics noise.}
	\label{fig:dynNoise}
\end{figure}

This type of noise can be considered as environmental noise that impacts the agents directly. Therefore dynamics of an agent under this uncertainty can be modeled as 

\begin{align}\label{agentUpdateDN}
\dot{x}_i=\sum_{j \neq i} {a_{ij}\big({x_j(t-\tau)-x_i(t-\tau)}\big)}+\xi_i(t),
\end{align}
where $\xi_i \sim \mathcal{N}(0,\sigma_i^2)$ for all $i \in \V$. Performance of a network with this type of  uncertainty structure was previously studied in \cite{siami2016fundamental,yaser2016delay} where dynamics of the network can be modeled by setting $B = I_{\N}$ in \eqref{eq:system}, i.e., 
\begin{align}\label{eq:systemDN}
\begin{aligned}
\dot{x}(t)~&= -L\,x(t-\tau)+\xi(t).
\end{aligned}
\end{align}
\noindent Figure \ref{fig:dynNoise} is a representation of dynamics of the network.

\begin{theorem}
	For consensus network \eqref{eq:systemDN}, centrality index of node $i$ is equal to
	\begin{align}\label{etaiDynamic}
	\eta_i = \frac{1}{2}\big[L^{\dagger}\cos(\tau L)\big(M_n - \sin(\tau L)\big)^{\dagger}\big]_{ii},
	\end{align}
	for all $i \in \V$.
\end{theorem}
\begin{pf}
	Since ${B} = I_{\N}$ and $\xi_i$'s are independent, using Theorem \ref{th:Cent12}, \eqref{etaiDynamic} can be followed.
\end{pf}
Taking derivative of $\eta_i$'s with respect to the delay parameter $\tau$, we obtain
\begin{align*}
\frac{\partial \eta_i}{\partial \tau} = \frac{1}{2}\big[\big(M_{\N}-\sin(\tau L)\big)^{\dagger}\big]_{ii},
\end{align*}
and since the function is not correlated to value of centrality index at $\tau = 0$, it hints us that centrality ordering can change as time-delay  increases and the ranking might get inverted. Later on, in Example \ref{ex1} we verify that this is the case and order inversion can happen.

\begin{theorem}
	For consensus network \eqref{eq:systemDN}, sensitivity coefficient of the interconnection between nodes $i$ and $j$ is equal to
	\begin{align*}
	\kappa_{e} = \frac{1}{2}r_{e}\Big(\big(L^2\big)^{\dagger}\big(\tau L - \cos(\tau L)\big)\big(M_n - \sin(\tau L)\big)^{\dagger}\Big),
	\end{align*}
	for all $i, j \in \V$.
\end{theorem}
\begin{pf}
	From definition of $\kappa_{e}$, taking derivative of $\rhoo(L;\tau)$ with respect to $w(e)$, we have
	\begin{align} 
	\kappa_{e} = \frac{1}{2}\Tr\Bigg[\NHS\NHS-\NHS L^{\dagger}E_{e}E_{e}^{\T}L^{\dagger}\cos(\tau L)\big(M_n - \sin(\tau L)\big)^{\dagger}\nonumber\\
	-\tau L^{\dagger}E_{e}E_{e}^{\T}\sin(\tau L)\big(M_n - \sin(\tau L)\big)^{\dagger}\nonumber\\
+\tau L^{\dagger}\cos^2(\tau L)E_{e}E_{e}^{\T}\Big(\NHS\big(M_n\NHS - \sin(\tau L)\big)^2\Big)^{\dagger}\NHS\Bigg],\label{centrality3Dynamic}
	\end{align}
	where $E_{e}$ is the corresponding column of the link $e$ in the incidence matrix of the graph.
	Then, since $L$, $\cos(\tau L)$, $\sin(\tau L)$, and $(M_n-\sin(\tau L))$ commute, and also trace is invariant under cyclic permutation, a proof follows by rearranging matrices in \eqref{centrality3Dynamic}.
\end{pf}

\subsection{Sensor Noise}
This type of noise as the name suggests, stems from uncertainties in the measurement of state of each agent measured by agent itself and eventually sent to other agents. As it is mentioned by \cite{siami2018centrality}, in an environment that is suffering from sensor noises and time-delay $\tau$, dynamics of each individual agent $i$ for all $i \in \V$ can be modeled as follows,
\begin{align}\label{agentUpdateSN}
\dot{x}_i=\sum_{j \neq i} {a_{ij}\Big({\big(x_j(t-\tau)+\xi_j(t)\big)-\big(x_i(t-\tau)+\xi_i(t)\big)}\Big)},
\end{align}
where $\xi \sim \mathcal N (0, \sigma_i^2)$ for $i \in \V$. The rationale behind such modeling is that each agent models its state by state of some other. 
Consequently, dynamics of network can be formulated by the input matrix ${B} = L$ as follows 
\begin{align}\label{eq:systemSN}
\begin{aligned}
\dot{x}(t)~&= -L\,x(t-\tau)+L\xi(t).
\end{aligned}
\end{align}

\begin{theorem}
	For consensus network \eqref{eq:systemDN}, centrality index of node $i$ is equal to
	\begin{align}\label{etaiSensor}
	\eta_i = \frac{1}{2}\big[L\cos(\tau L)\big(M_n - \sin(\tau L)\big)^{\dagger}\big]_{ii},
	\end{align}
	for all $i \in \V$.
\end{theorem}
\begin{pf}
	Since ${B} = L$ and $\xi_i$'s are independent, using Theorem \ref{th:Cent12}, equation \eqref{etaiSensor} can be followed.
\end{pf}

\begin{theorem}
	For consensus network \eqref{eq:systemDN}, sensitivity coefficient of the interconnection between nodes $i$ and $j$ is equal to
	\begin{align}\label{kappaiDynamic}
	\kappa_{e} = \frac{1}{2}r_{e}\Big(\big(\tau L + \cos(\tau L)\big)\big(M_n - \sin(\tau L)\big)^{\dagger}\Big),
	\end{align}
	for all $i, j \in \V$.
\end{theorem}
\begin{pf}
	From definition of $\kappa_{e}$, taking derivative of $\rhoo(L;\tau)$ with respect to $w(e)$, we have
	\begin{align} 
	\kappa_{e} = \frac{1}{2}\Tr\Bigg[E_{e}E_{e}^{\T}\cos(\tau L)\big(M_n - \sin(\tau L)\big)^{\dagger}\nonumber\\
	-\tau E_{e}E_{e}^{\T}\sin(\tau L)\big(M_n - \sin(\tau L)\big)^{\dagger}\nonumber\\
	+\tau \cos^2(\tau L)E_{e}E_{e}^{\T}\Big(\NHS\big(M_n\NHS - \sin(\tau L)\big)^2\Big)^{\dagger}\Bigg],\label{centrality3Noise}
	\end{align}
	where $E_{e}$ is the corresponding column of the link $e$ in the incidence matrix of the graph.
	Then, since $L$, $\cos(\tau L)$, $\sin(\tau L)$, and $(M_n-\sin(\tau L))$ commute, and also trace is invariant under cyclic permutation, a proof follows by rearranging matrices in \eqref{centrality3Noise}.
\end{pf}

\begin{figure}
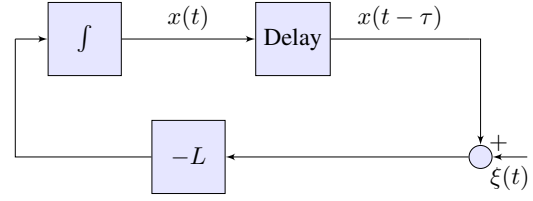

	\centering
	\includestandalone[width=0.42\textwidth]{snsrNoise}
	\caption{Block diagram of consensus network \eqref{eq:system} in presence of sensor noise.}
	\label{fig:snsrNoise}
\end{figure}

\subsection{Receiver Noise}
This type of uncertainty emerges when there exist noise on the receiver node. In other words, when agents $i$ is receiving $x_j(t-\tau)+\xi_i(t)$ as state the of agent $j$ whereas in absence of the disturbance, it would have received $x_j(t-\tau)$ as state of agent $j$. Consequently, when there exists such receiver noise in the system, the update law for dynamics is given by
\begin{align}\label{agentUpdateRN}
\dot{x}_i=\sum_{j \neq i} {a_{ij}\Big({\big(x_j(t-\tau)+\xi_i(t)\big)-x_i(t-\tau)}\Big)},
\end{align}
for all agents $i \in \V$, where $\xi_i \sim \mathcal{N}(0,\sigma_i^2)$.
This dynamics can be cast in the form of the consensus dynamics \eqref{eq:system}, where $B = \Delta$ which is the diagonal matrix of degrees of the nodes. The following theorem, discuss centrality of the agents in such network.

\begin{theorem}
	For consensus network with update law \eqref{agentUpdateRN}, centrality index of agent $i$ is equal to
	\begin{align}\label{etaiReceiver}
	\eta_i = \frac{1}{2}\big[\Delta^2 L^{\dagger}\cos(\tau L)\big(M_n - \sin(\tau L)\big)^{\dagger}\big]_{ii},
	\end{align}
	for all $i \in \V$.
\end{theorem}
\begin{pf}
	Since consensus network \eqref{agentUpdateRN} is a special case of \eqref{eq:system} with $B = \Delta$ and $\xi_i$'s are independent, using Theorem \ref{th:Cent12}, equation \eqref{etaiReceiver} follows immediately.
\end{pf}

\subsection{Emitter Noise}
This type of noise can be generated by signal emitter of the agents and therefore as a result cause an uncertainty in signals that are received by neighboring agents. Thus, when agent $j$ sends its state $x_j(t-\tau)$ to the node $i$, what node $i$ receives is $x_j(t-\tau) +\xi_j(t)$. As a result, dynamics of each agent $i$ can be modeled by

\begin{align}\label{agentUpdateEN}
\dot{x}_i=\sum_{j \neq i} {a_{ij}\Big({\big(x_j(t-\tau)+\xi_j(t)\big)-x_i(t-\tau)}\Big)},
\end{align}
for all agents $i \in \V$, where $\xi_i \sim \mathcal{N}(0,\sigma_i^2)$.
This dynamics can be cast in the form of the consensus dynamics \eqref{eq:system}, with adjacency matrix of the underlying graph as the matrix $B$. 

\begin{theorem}
	For consensus network with update law \eqref{agentUpdateEN}, centrality index of agent $i$ is equal to
	\begin{align*}
	\eta_i = \frac{1}{2}\big[\big(\Delta^2 L^{\dagger}-\Delta +L\big)\cos(\tau L)\big(M_n - \sin(\tau L)\big)^{\dagger}\big]_{ii},
	\end{align*}
	for all $i \in \V$.
\end{theorem}
\begin{pf}
	Observe that adjacency matrix $A = \Delta - L$ and consensus network \eqref{agentUpdateRN} is a special case of \eqref{eq:system} with adjacency matrix, $A$, of the graph as the input matrix, we have $B = A = \Delta - L$. Also, since $\xi_i$'s are independent, using Theorem \ref{th:Cent12},
	we obtain 
	\begin{align*}
	\eta_i = \frac{1}{2}\big[\big(\Delta - L\big)L^{\dagger}\cos(\tau L)\big(M_n - \sin(\tau L)\big)^{\dagger}\big(\Delta - L\big)\big]_{ii}.
	\end{align*}
	Finally, trace operator is invariant under cyclic permutation which yields the result.
\end{pf}

\section{Link Associated Disturbances}
In this section with discuss the noises that affect the links between agents and their associated centrality indices. 
\begin{figure}
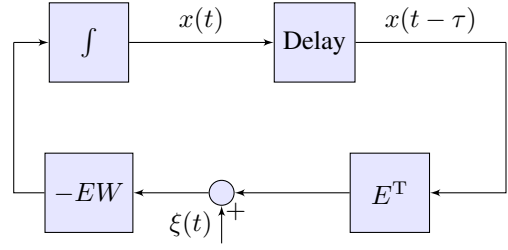

	\centering
	\includestandalone[width=0.40\textwidth]{comNoise}
	\caption{Block diagram of consensus network \eqref{eq:system} in presence of communication noise.}
	\label{fig:comNoise}
\end{figure}
\subsection{Communication Channel Noise}

This type of noise may arise because of signal distortion in a communication channel between two agents in the network. We assumed that each  communication channel suffers from a Gaussian noise  $\xi_e \sim \mathcal {N}(0, \sigma_e^2)$ for all $e \in \EE$. Under this assumption, if agents $i$ and $j$ are communicating through the channel $e = \{i,j\}$, agent $i$, receives $x_j(t-\tau) +\xi_e(t)$, instead of $x_j(t-\tau)$ while agent $j$, receives $x_i(t-\tau) -\xi_e(t)$ rather than $x_i(t-\tau)$. In other words, the relative state of agents on head end of the communication channel is modified by $\xi_e(t)$ and the tail end is adjusted by $\xi_e(t)$. As a result, we obtain oriented incidence matrix $E$, however, we note that since the graph is undirected and Gaussian distribution is symmetric, choice of the head and tail ends for a link does not affect dynamics of the network. With this being the case, each agent uses the following update law

\begin{align}\label{agentUpdateCN}
\dot{x}_i\,=\sum_{e = \{i,j\} \in \EE} {a_{ij}\big({x_j(t-\tau)-x_i(t-\tau)+\xi_e(t)}\big)},
\end{align}
and since the noise on each link is independent of the other, dynamics of the network can be formulated in the form of \eqref{eq:system}, where $B = E W$.
Figure \ref{fig:comNoise} illustrates structure of a network with this type of uncertainty.
\begin{theorem}
	For consensus network with update algorithm \eqref{agentUpdateCN}, value of centrality for link $e = \{i,j\}$ is 
	\begin{align*}
	\nu_e =\frac{1}{2} a_{e}^2r_{e}\Big(L\cos(\tau L)^{-1}\big(M_n - \sin(\tau L)\big)\Big).
	\end{align*}
\end{theorem}
\vspace{0.2cm}
{
	\begin{pf}
		Using the same technique used for proof of Theorem  \ref{th:Cent12}, matrix $\hat{B} = E W \diag([\sigma_1^2,\dots,\sigma_{\mid \EE \mid}^2]^{\T})$ and consequently, we obtain the performance measure as
		\begin{align}
		&\rhoo(L;\tau) = \frac{1}{2}\Tr\NHS\NHS\big[\hat{B}^{T}L^{\dagger}\cos(\tau L)\big(M_n - \sin(\tau L)\big)^{\dagger}\hat{B}\big]\nonumber\\
		\NHS\NHS\NHS&\NHS\NHS\NHS=\NHS\frac{1}{2}\NHS\Tr\NHS\big[\NHS\diag([\sigma_1^2,\NHS\dots\NHS,\NHS\sigma_{\mid \EE \mid}^2]^{\T})\NHS W^2 \NHS E^{T}\NHS L^{\dagger}\NHS\cos(\tau\NHS L)\big(M_n \NHS\NHS-\NHS \NHS \sin(\tau L)\big)^{\dagger}\NHS E\NHS\big]\nonumber\\
		\NHS\NHS\NHS&\NHS\NHS\NHS=\frac{1}{2}\sum_{e=\{i,j\} \in \EE}\sigma_e^2 a_{ij}^2r_e\big(L\cos(\tau L)^{-1}\big(M_n - \sin(\tau L)\big)\big)\label{rhoocomm}
		\end{align}
		A proof follows by taking derivative of \eqref{rhoocomm} with respect to $\sigma_e^2$ for all $e \in \EE$.
\end{pf}}
\subsection{Measurement Noise}

\begin{figure}
\centering
	\includestandalone[width=0.35\textwidth]{measNoise}
	\caption{Block diagram of consensus network \eqref{eq:system} in presence of measurement noise.}
	\label{fig:measNoise}
\end{figure}

\begin{figure}[t]
	\centering
	\includegraphics[width=0.27\textwidth]{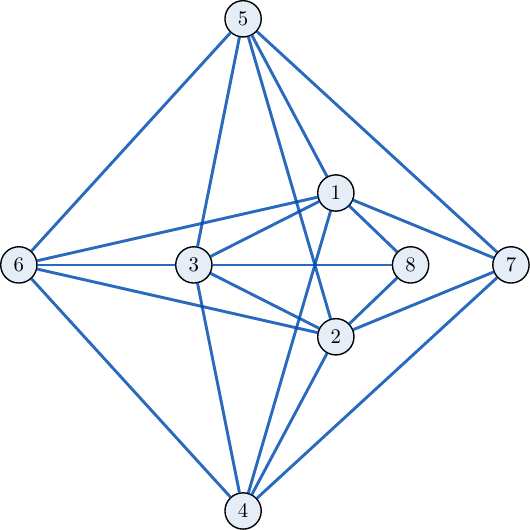}
	\caption{Graph of the example \ref{ex1} with 8 nodes and 20 links.}
	\label{fig_3}
\end{figure}

\begin{figure}[t]
	\centering
	\includegraphics[width=0.82\linewidth,trim={0.5cm 0 0.5cm 0.5cm}]{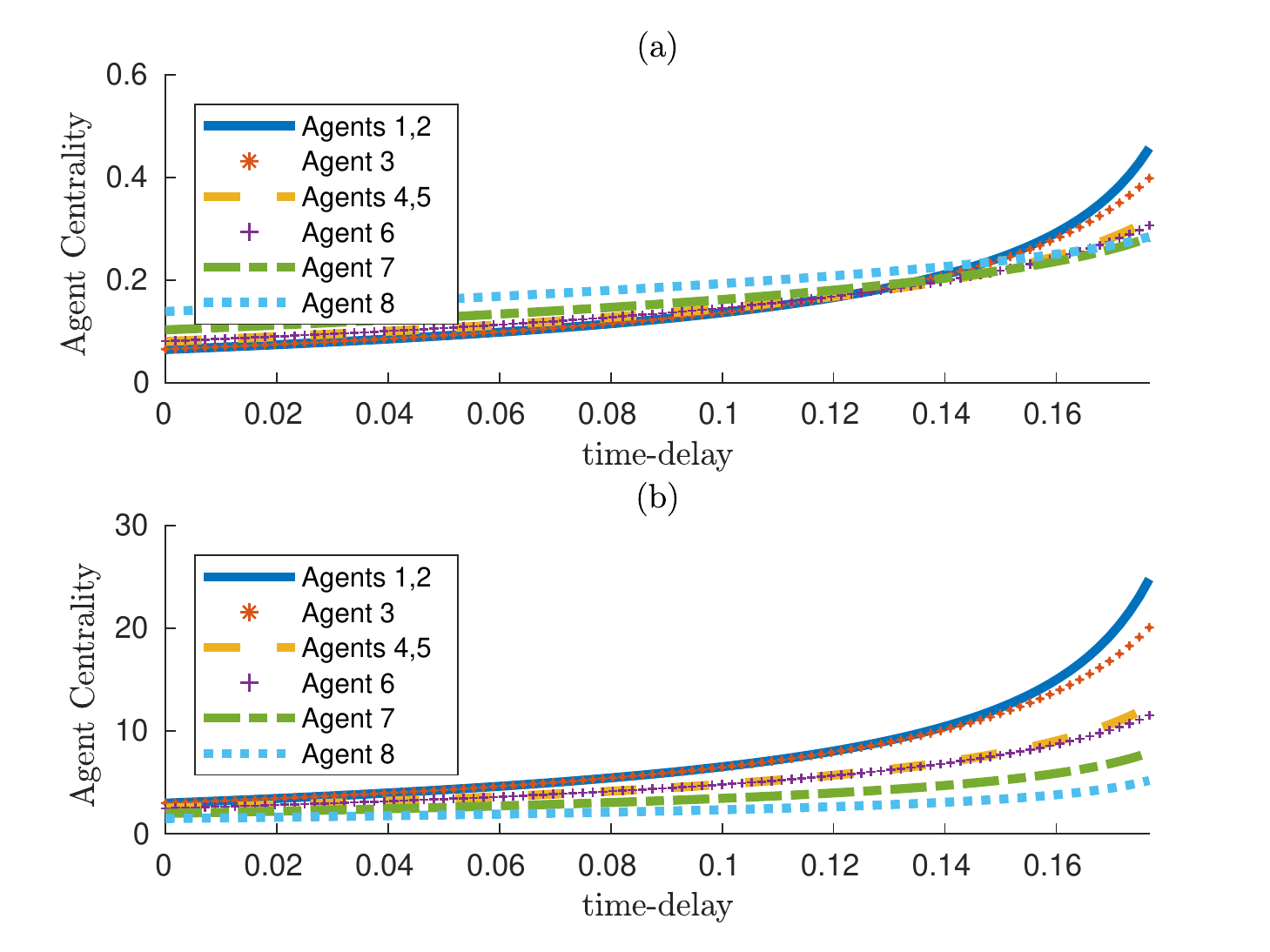}
	\caption{{ Centrality index as a function of time-delay }(a) Agent centrality with dynamics noise in example \ref{ex1}, (b) Agent centrality with sensor noise in example \ref{ex1}.}
	\label{fig_1}
\end{figure}

\begin{figure}[t]
	\centering
	\includegraphics[width=0.82\linewidth,trim={0.5cm 0 0.5cm 0.5cm}]{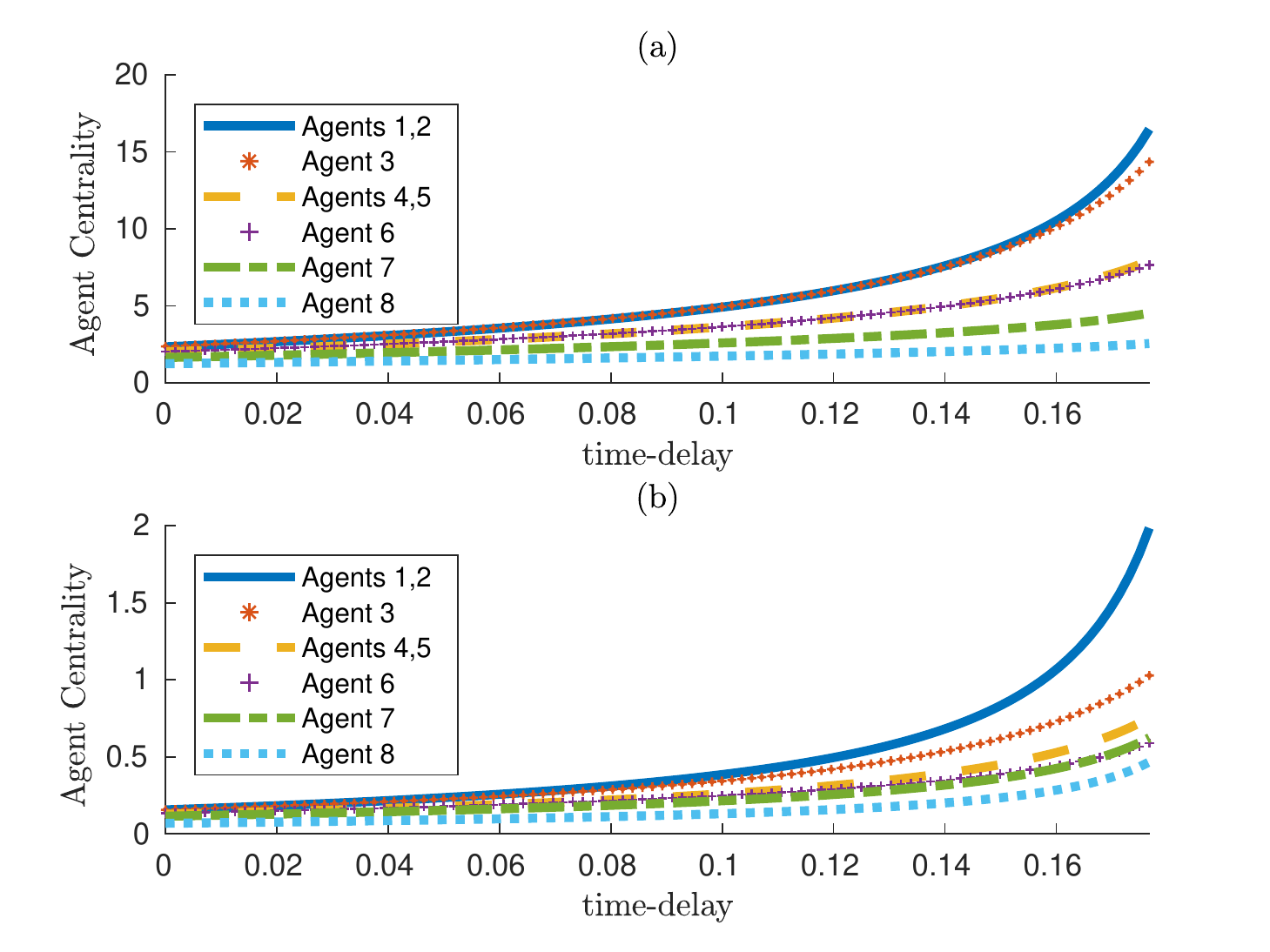}
	\caption{{ Centrality index as a function of time-delay }(a) Agent centrality with receiver noise in example \ref{ex1},(b) Agent centrality with emitter noise in example \ref{ex1}.}
	\label{fig_2}
\end{figure}

{ This type of noise is used to mimic the
	effect of measurement noise that occurs in practice (see \cite{siami2018centrality} for details). We can use the two-port representation of linear consensus network as described in \cite{Zelazo:2011,siami2018centrality}, then it follows that
	\begin{eqnarray}\label{updatem1}
	\dot{x}(t) & = & u(t), \\
	z(t) & = & W E^{\rm T} x(t) + \xi(t),\label{updatem2}
	\end{eqnarray}
	where $\xi(t)= \left[~\xi_{e_1},~ \ldots~,~\xi_{e_{\mid \EE \mid}} ~\right]^{\rm T}$ is the vector of input noise, $\xi_{e}(t) \sim N(0,\sigma_{e}^2)$ for all $e \in \mathcal{E}$, $E$ is the signed vertex-edge
	incidence matrix of the graph, and the internal feedback control law is given by 
	\begin{equation*}
	u(t) = - E z(t). 
	\end{equation*}
	By direct calculations, one can verify that dynamics of the network can be formulated in the form of \eqref{eq:system}, where ${B = - E}$.
	Figure \ref{fig:measNoise} depicts a representation of this linear consensus network with measurement noises.
} 

\begin{theorem}
	For consensus network with update algorithm \eqref{updatem1}, \eqref{updatem2} value of centrality for link $e = \{i,j\}$ is 
	\begin{align*}
	\nu_e =\frac{1}{2} r_e\Big(L\cos(\tau L)^{-1}\big(M_n - \sin(\tau L)\big)\Big).
	\end{align*}
\end{theorem}
\vspace{0.2cm}
{
	\begin{pf}
		Using the same technique used for proof of Theorem  \ref{th:Cent12}, matrix $\hat{B} = E \diag([\sigma_1^2,\dots,\sigma_{\mid \EE \mid}^2]^{\T})$ and thus, we can obtain the performance measure as
		\begin{align}
		&\rhoo(L;\tau) = \frac{1}{2}\Tr\NHS\big[\hat{B}^{T}L^{\dagger}\cos(\tau L)\big(M_n - \sin(\tau L)\big)^{\dagger}\hat{B}\big]\nonumber\\
		\NHS\NHS\NHS&\NHS\NHS\NHS=\NHS\NHS\frac{1}{2}\Tr\big[\NHS\diag([\sigma_1^2,\NHS\dots\NHS,\NHS\sigma_{m}^2]^{\T})\NHS  E^{T}\NHS L^{\dagger}\NHS\cos(\tau\NHS L)\big(M_n \NHS\NHS- \sin(\tau L)\big)^{\dagger}E\big]\nonumber\\
		\NHS\NHS\NHS&\NHS\NHS\NHS=\frac{1}{2}\sum_{e=\{i,j\} \in \EE}\sigma_e^2 r_e\big(L\cos(\tau L)^{-1}\big(M_n - \sin(\tau L)\big)\big)\label{rhoomeas}
		\end{align}
		A proof follows by taking derivative of \eqref{rhoomeas} with respect to $\sigma_e^2$ for all $e \in \EE$.
\end{pf}}

\section{Order of Precedence and Effect of Connectivity}\label{sec:ranking}
One natural way to characterize the effect of noise structures on a dynamic network is by ordering the agents' or links' centrality indices,  \cite{siami2018centrality}. 
We introduce the term \textit{precedence} when we refer to order of node centrality and the term \textit{ranking} when we refer to order link centrality. 

\begin{definition}
	In a network with $n$ agents, we say agent $i$ is higher in the {order of precedence} than agent $j$, if $\eta_i > \eta_j$. Moreover, for links $e_i$ and $e_j$, we say link $e_i$ has a higher rank than link $e_j$, if $\nu_{e_i} > \nu_{e_j}$.	
\end{definition}

 Theorem below, discusses the effect of uniform scaling of the connectivity across the network.

\begin{theorem}\label{thm:uniformScaling}
	In the absence of time-delay, for all types of uncertainties, order of precedence of the agents and ranking of the links is invariant with respect to uniform scaling of the weight of all links.
\end{theorem}
{
	\begin{pf}
		In the absence of time-delay, $\cos(\tau L) = I_{\N}$, and $\sin(\tau L) = 0_{\N\times \N}$. In addition, scaling all the weights matrices by $\alpha>0$, scales $\Delta$, $L$ by $\alpha$, and scales $L^{\dagger}$ by $1/\alpha$. Thus, scaling the weights by $\alpha$, scales agent centrality indices with dynamics noise and link centrality measures by $1/\alpha$. Similarly,  agent centrality indices with receiver noise, sensor noise, and emitter noise will be scaled by $\alpha$.
\end{pf}}

According to Theorem \ref{thm:uniformScaling}, in a synchronous consensus network, precedence and ranking remains invariant under uniform scaling of the coupling weights. This is not the case however when time-delay is present. Later in Example \ref{ex:uniform_connectivity}, we see that in the presence of time-delay, uniform scaling of the weights can indeed change ranking among the agents. 

The interplay between network connectivity, time-delay and ordering over nodes/links in a consensus network is quite perplexed. In the next result we establish an asymptotic relation between order of precedence and ranking of the links in the presence and absence of time-delay.

\begin{theorem}\label{thm:uniformScaling2}
	For a network with Laplacian matrix $L$ and a specific type of uncertainty, without loss of generality, assume that the agents are labeled based on their order of precedence in the absence of time-delay, i.e., agent $i$ precedes agent $j$ in rank, if and only if $i < j$. Similarly, assume that links are labeled, based on their rank in the absence of time-delay. Then, in the presence of time-delay $\tau$, there exist a positive scalar $\alpha$, such that a network with the Laplacian matrix $\alpha L$, in which agent $i$ achieves rank $i$ for all $i \in \{1,2,\dots,\N\}$ and link $e$ achieves rank $e$, for all $e \in \{1,2,\dots,\mid\EE\mid\}$. In other words, in the presence of time-delay, scaling all links by a small enough $\alpha$, provides the same ranking to the delay-free case with the same type of uncertainty.
\end{theorem}
\begin{pf}
	If we let $\alpha$ converge to zero, then $\cos(\tau \alpha L) $ converges to $ I_{\N}$ and  $\sin(\tau \alpha L) $ approaches $ 0_{\N \times \N}$ and thus, $\Big(M_{\N}-\sin(\tau \alpha L)\Big)^{\dagger} $ approaches $ M_{\N}^{\dagger} = M_{\N}$. Thus, the centrality index converges to that of a network in the absence of time-delay with Laplacian matrix $\alpha L$. The rest of the proof follows by applying Theorem \ref{thm:uniformScaling} since the centrality ranking of nodes in the absence of time-delay is invariant with respect to scaling.
\end{pf}

Theorem \ref{thm:uniformScaling2} relies on the continuous dependence of the centrality indices with respect to coupling weights. It explains that for given time-delay $\tau>0$ there is a small enough uniform scaling parameter that can match the effect of noise on the network when time-delay is not present. Such scaling parameter $\alpha$ is in fact independent of network structure $L$.

\begin{figure}[t]
	\centering
	\includegraphics[width=0.29\textwidth]{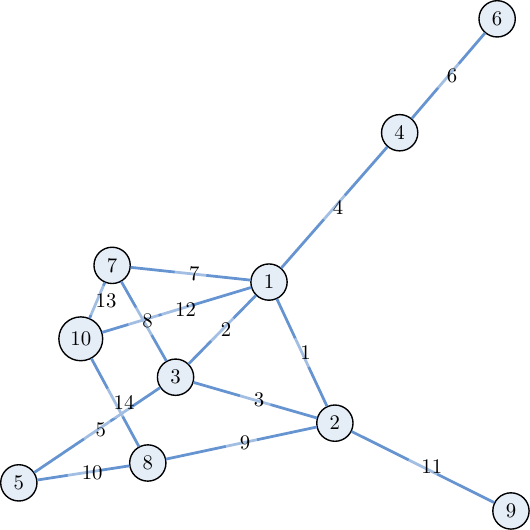}
	\caption{\footnotesize{Erd\H{o}s-R\'enyi graph with $ n = 10$ nodes and probability $p = 0.3$.}}     \label{Fig:exampleGraph_2}
\end{figure}

\section{Numerical Examples}
\begin{example}\label{ex1}
	In this example, we consider a randomly generated network with $8$ agents and $20$ unweighted links illustrated in Figure \ref{fig_3}. We study centrality indices of agents in presence of 4 different uncertainty structures for different amount of time-delay. We note that as expected, all indices increase with the time-delay. In addition, we observe that since agents  { 1 and 2} share same set of neighbors, i.e., there is an automorphism that maps them to one another, all their centrality indices are equal. Similarly, all indices of agents { 4 and 5}, are equal. { In this example, agent labeling is based on the value of centrality index in a consensus network with dynamics noise in absence of time-delay. More specifically, agents with greater centrality index in presence of dynamics noise, have greater label.} Interestingly, in Figures \ref{fig_1} and \ref{fig_2} we observe that centrality rankings are not invariant with respect to time-delay. { Also, another noteworthy observation is that although centrality rankings for different noise structures do not match in absence of time-delay \cite{siami2018centrality}, they are very similar to each other as time-delay increases. Our intuition behind this phenomenon is that as time-delay increases, eigenvectors of larger eigenvalues of the Laplacian matrix play a major role especially as $\tau \to \frac{\pi}{2\lambda_n}$}.
	
\end{example}

\begin{example}\label{ex:uniform_connectivity}

\begin{figure}[t]
	\centering
	\includegraphics[width=0.75\linewidth]{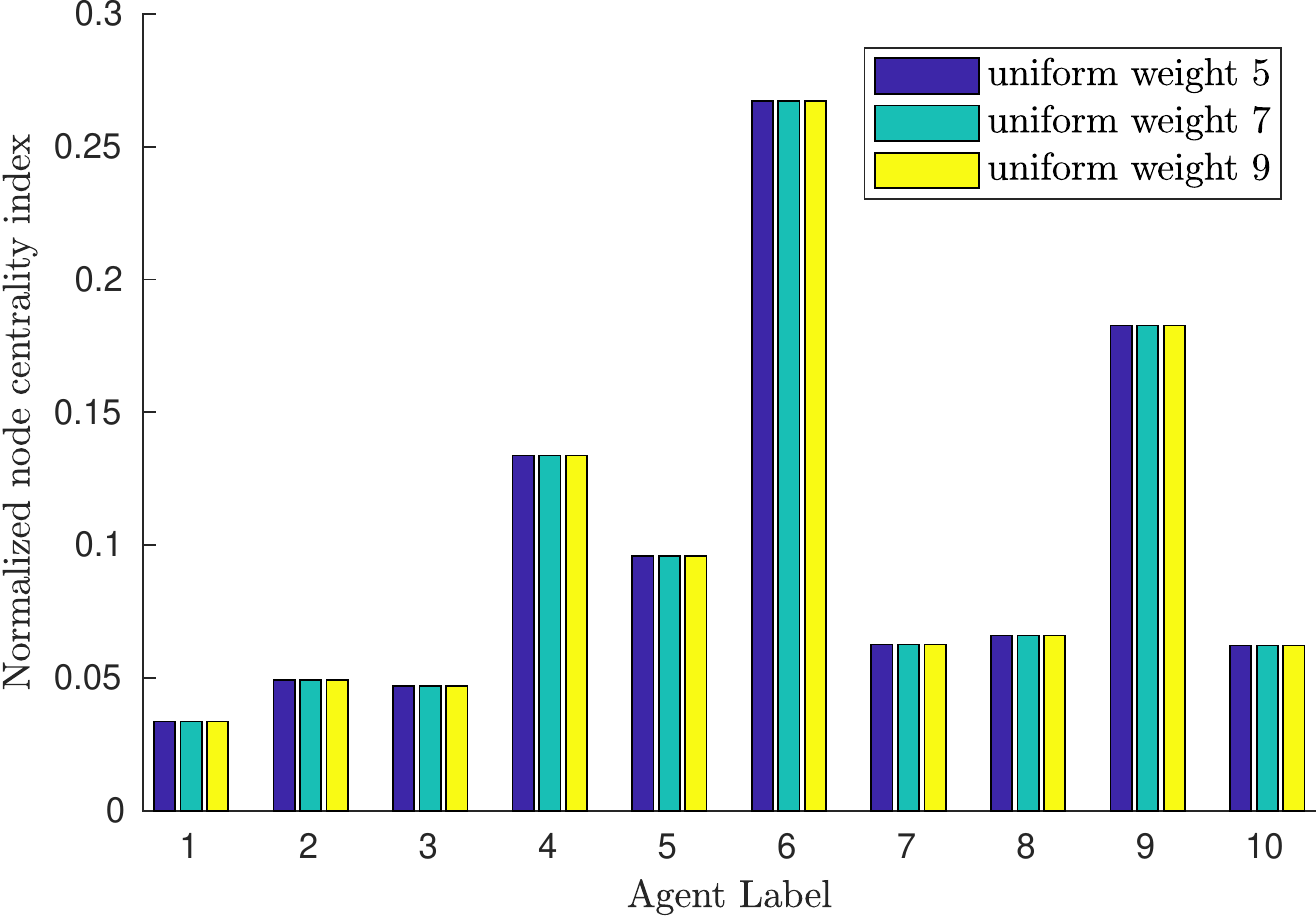}
	\caption{Normalized agent centrality index with dynamics noise in the absence of time-delay, is invariant with respect to uniform scaling of the weights.}
	\label{Fig:exam2_nodelay_node}
\end{figure}
\vspace{2cm} 
\begin{figure}[t]
	\centering
	\includegraphics[width=0.75\linewidth]{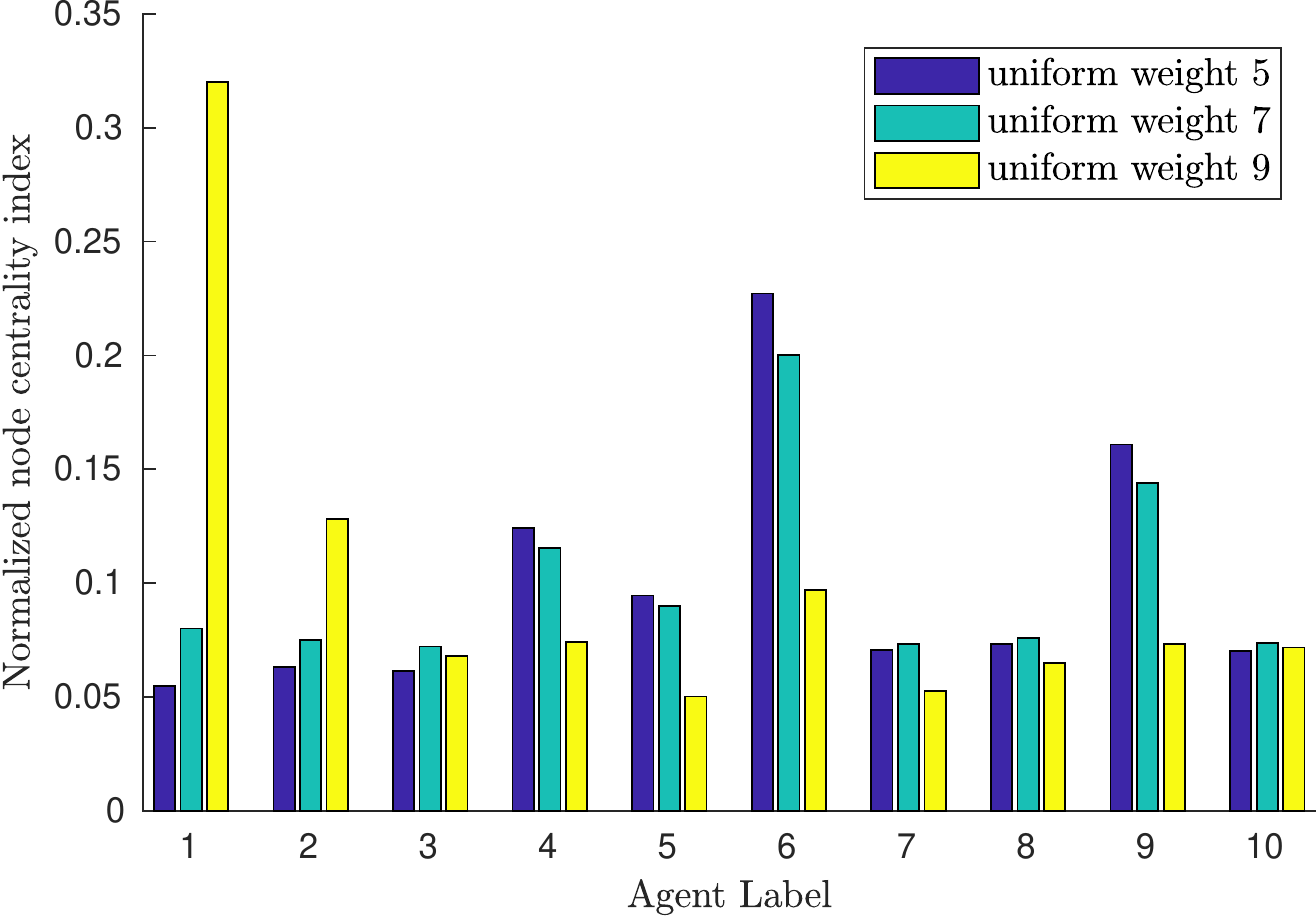}
	\caption{Normalized agent centrality index with additive noise in the presence of time-delay.}
	\label{Fig:exam2_wdelay_node}
\end{figure}

	In order to study effect of time-delay and connectivity on the centrality ranking of agents and links, consider the Erd\H{o}s-R\'enyi graph on $ n = 10$ agents and each link is included with probability $p = 0.3$, depicted in Figure \ref{Fig:exampleGraph_2}. First, we study centrality of the agents in the absence of time-delay. We consider three different uniform weights of $5, 7,$ and $9$ across the network. From Theorem \ref{thm:uniformScaling}, we expect the ranking of the agents to be unaffected by this change in connectivity. Thus, measuring the agent centrality in presence of the dynamics noise, normalized (i.e., divided by  the sum of all indices) value of centrality indices, in Figure \ref{Fig:exam2_nodelay_node} is constant for all three weights. This means that the ranking is not changed by increasing the connectivity. On the other hand, when time-delay $\tau = 0.268$, increasing connectivity from $5$ to $7$ and $9$, changes the ranking of the agents. For example, in the presence of time-delay, agent $6$ and agent $1$ has the highest and lowest order of precedence, respectively, when the uniform weight of the couplings is $5$. On the contrary, when the uniform weights are increased to $9$, agent $1$ which was the lowest agent in the ranking, achieves the highest ranking among the agents and agent $6$ which had the highest rank, is demoted to the third place. Furthermore, we can observe the relative change of the centrality index and the ranking for other agents in Figure \ref{Fig:exam2_wdelay_node}.

	\begin{figure}[t]
		\centering
		\includegraphics[width=0.75\linewidth]{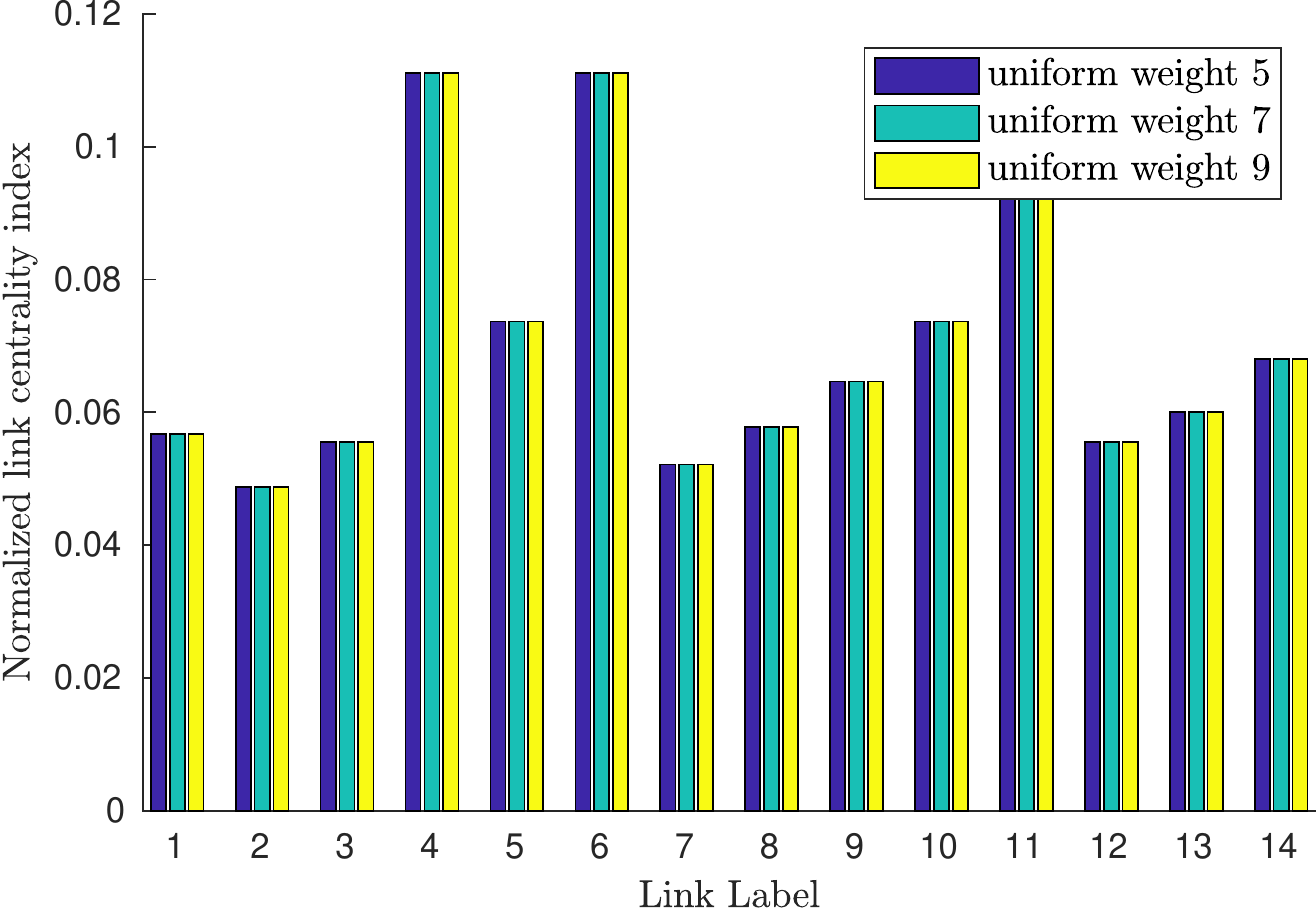}
		\caption{Normalized link centrality index with dynamics noise in the absence of time-delay, is invariant with respect to uniform scaling of the weights.}
		\label{Fig:exam2_nodelay_link}
	\end{figure}
	
	\begin{figure}[t]
		\centering
		\includegraphics[width=0.75\linewidth]{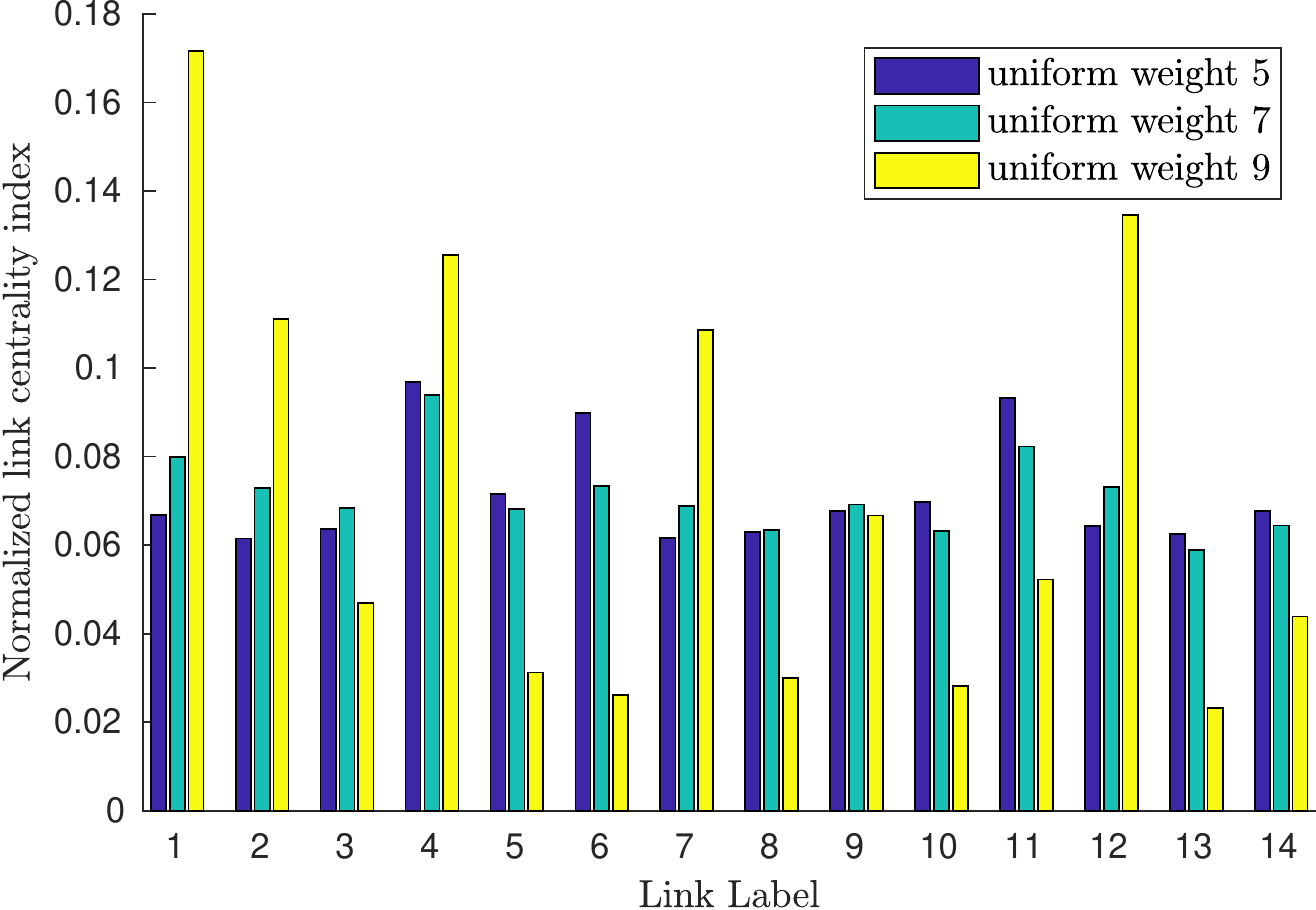}
		\caption{Normalized link centrality index with additive noise in the presence of time-delay.}
		\label{Fig:exam2_wdelay_link}
	\end{figure}
	
\end{example}

\begin{example}\label{ex:facebook}
	In this example we consider a dataset of Facebook users \cite{snapnets} and we analyze the effect of time-delay on link centrality indices and rankings. To do so, we find top 10 ranked links in the network both in the presence and in the absence of time-delay. Our observation is that by adding time-delay, the links close to the agent with highest degree become the links with higher rank.

	\begin{figure*}[!htb]
		\begin{minipage}{0.4\textwidth}
			\centering
			\includegraphics[width=1.1\linewidth
			,trim={3.205cm 1.35cm 0 0},clip
			]{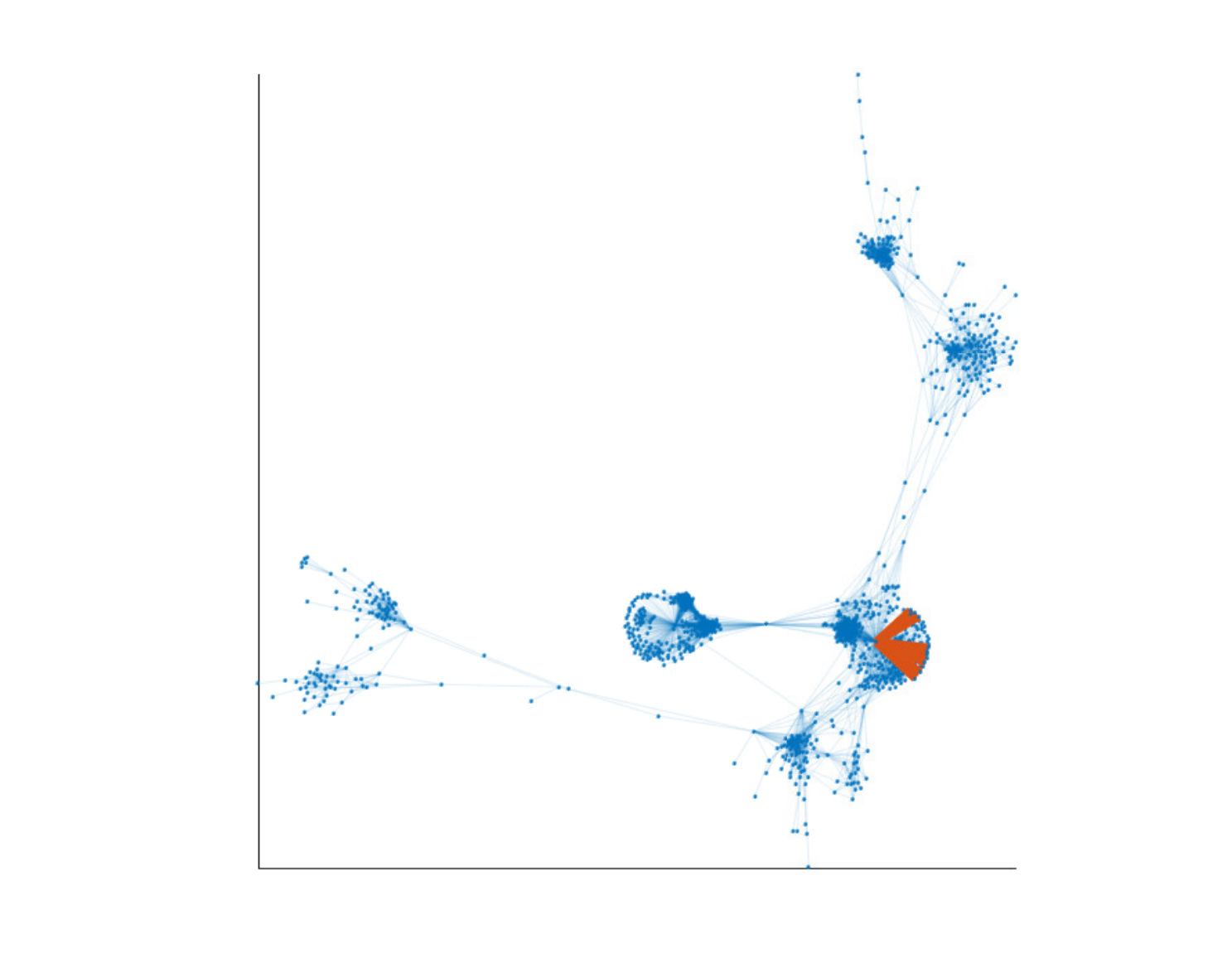}
			\caption{The graph of Example \ref{ex:facebook}, the 10 red links are those with highest centrality index in the presence of time-delay and they are in the vicinity of the agent with the largest degree. }
			\label{Fig:exam_facebook_delayed}
		\end{minipage}
		\hfill
		\begin{minipage}{0.4\textwidth}
			\centering
			\includegraphics[width=1.1\linewidth
			,trim={3.356cm 1.25cm 0 0},clip
			] {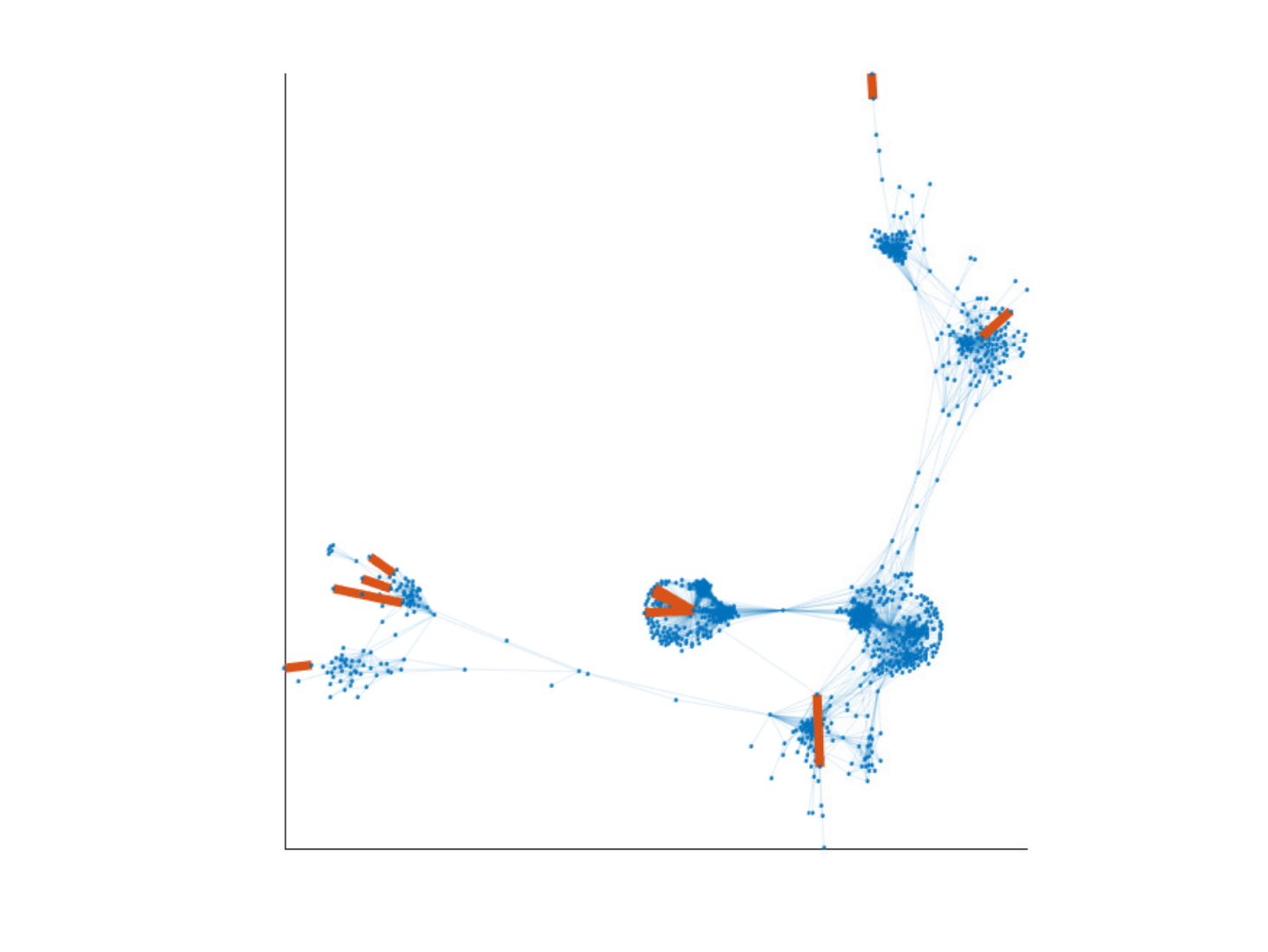}
			\caption{The graph of Example \ref{ex:facebook}, the 10 red links are those with highest centrality index in the absence of time-delay and they are spread far from the agents with the largest degree. }
			\label{Fig:exam_facebook_delayfree}
		\end{minipage}
	\end{figure*}
	
\end{example}

{\section{Discussions}\label{sec:discussions}
	{
		\subsection{Computational Complexity}
		The first step in computing the centrality indices for a agent or a link, requires finding $L^{\dagger} = (L+\frac{1}{\N}J_{\N})^{-1}-\frac{1}{n}J_{\N}$ which requires $\mathcal{O}({\N}^3)$ arithmetic operations. Then we need to find $\cos(\tau L)$ and $\sin(\tau L)$ using pad\'e approximation. Computational complexity of this step is $\mathcal{O}({\N}^3)$ as well \cite{hargreaves2005efficient}. Quite similarly, computational complexity of finding finding $\big(M_n - \sin(\tau L)\big)^{\dagger}$ is $\mathcal{O}({\N}^3)$. Then we need to do up to four matrix multiplications between dense matrices which needs $\mathcal{O}({\N}^3)$. Summing up all the required operations, results in $\mathcal{O}(\N^3)$ computational complexity.
		
		\subsection{Networks with higher-order dynamics}
		The results of this manuscript can be extended to study influence of agents in a network with higher-order dynamics or multi-layered structure \cite{alemzadeh2018influence}. We discuss the extension of the results to a class of networks with second-order dynamics which is a widely used model for studying platoon of cars \cite{ren2008consensus,yu2010some,bamieh2012coherence}. Even though we discuss the centrality of the agents in the presences of dynamics noise, centrality of the agents and links can be driven for other types of uncertainty as well. For this class of networks, dynamics of the agent $i$ can be written as 
		
		\begin{align}
		\dot{x}_i(t) &= v_i,\\
		\dot{v}_i (t)&= \sum_{j \neq i}{a_{ij}\big({x_j(t-\tau)-x_i(t-\tau)}\big)} \nonumber\\
		&+b\sum_{j \neq i}  {a_{ij}\big({v_j(t-\tau)-v_i(t-\tau)}\big)} + \xi_i(t),\label{agents_with_input}
		\end{align}
		where $x_i$ is the position of agent $i$ and $v_i$ is the velocity of agent $i$.
		
		Thus, the network that we address in this section is the following second-order consensus network with $\N$ agents and uniform time-delay $\tau$ and underlying graph Laplacian $L$
		\begin{align}
		\begin{aligned}
		\label{eq:system_SOC}
		\dot{x}(t)~&= v(t),\\
		\dot{v}(t)~&= -Lx(t-\tau)-bL v(t-\tau) + \xi(t),\\
		y(t)~&=~ M_{\N}x(t),
		\end{aligned}
		\end{align}
		where $x = [x_1,  \ldots,  x_{\N}]^{\T}$ and $v = [v_1,  \ldots,  v_{\N}]^{\T}$ are states and $y = [y_1,  \ldots,  y_{\N}]^{\T}$ is the output.
		
		\begin{theorem}
			For consensus network \eqref{eq:system_SOC}, centrality index of agent $i$ is equal to
			\begin{align}\label{platoon_cent}
			\eta_i = \sum_{j=1}^{\N}Q_{ij}^2f(\lambda_i,\tau,b).
			\end{align}
			where function $Q_{ij}$ is the $i^{\mathrm {th}}$ element of the the normalized eigenvector corresponding to $\lambda_j$,  $$f(\lambda_i,\tau,b)= \int_{-\infty}^{+\infty}\frac{1}{2\pi}\frac{ d\omega}{h(\lambda_i,\tau,b,\omega)},$$ and
			\begin{align*}
			h(\lambda_i,\tau,b,\omega) = &{(\lambda _{i}}-\omega ^2\cos(\omega \tau))^2+\omega^2(b\lambda_i-\omega\sin(\omega\tau))^2.
			\end{align*}
			
		\end{theorem}

		\begin{corollary}
			In the absence of time-delay, in the second-order network \eqref{eq:system_SOC} with dynamics noise, centrality of the agents is simplified to 
			\begin{align*}
			\eta_i = \frac{1}{2b}\Big[\big(L^2\big)^{\dagger}\Big]_{ii}.
			\end{align*}
			
		\end{corollary} 
		\subsection{Designing robust networks}
		As it was discussed in Theorem \ref{th:Cent12}, centrality indices studied in this manuscript have a direct correlation with networks $\HH_2$-norm performance measure, and in fact, the performance of the network is a linear combination of the centrality indices. Thus, improving the indices can improve the performance of the network as well. However, adjusting centrality index of a agent might counter-effect the index of another agent. Thus, designing a network to improve centrality index of more than one agent (or link) is inherently a multi-objective optimization problem. We discuss the design problem in a network with dynamics noise but generalization for other types of uncertainty is possible. Scalarization \cite{marler2004survey} is a well-known class of approaches for solving multi-objective problems. For example we may consider a weighted sum of the centrality indices. From \ref{th:Cent12}, minimizing th weighted sum is equivalent to optimizing performance of the network. 
		Another viable approach for scalarization of the multi-objective problem is considering the $L_{\infty}$-norm of the vector of centrality indices. Since centrality of the agents are positively correlated with variability of the state of agent with respect to average of the state of all agents, minimizing the  $L_{\infty}$-norm of the centrality index of all agents improve the worst-case centrality in the network. If weight of the existing links in the network be a decision variable for improving the robustness of the network, then we can write the optimization problem in the following form
		
		\begin{flalign}\label{prob:opt1}
		\begin{aligned}
		& \underset{w(e),\forall e\in \EE}{\text{minimize}}
		& & \underset{i}{\text{maximize }} \eta_i \\
		& \text{subject to:}
		& & L = E_{e} E_{e}^{\T} w(e),\\
		& & & \eta_i\! =\! \frac{1}{2}\big[L^{\dagger}\cos(\tau L)\big(M_n - \sin(\tau L)\big)^{\dagger}\big]_{ii},\; \forall i\in \V 
		\end{aligned}
		\end{flalign}
		
		In the case that the disturbance on the system comes from an adversarial source with a \text{limited power}, the attacker aims at deteriorating the performance, while the goal is to design a network that is robust against the worst type of the attack. Thus, we need to solve the optimization problem 
		
		\begin{flalign}\label{prob:opt2}
		\begin{aligned}
		& \underset{w(e),\forall e\in \EE}{\text{minimize}}
		& & \underset{\sigma_i ,\forall i\in \V}{\text{maximize }} \rhoo(L;\tau) \\
		& \text{subject to:}
		& & L = E_{e} E_{e}^{\T} w(e)\\
		& & & \sum_{i=1}^{\N} \sigma_i^2 = \N.
		\end{aligned}&&
		\end{flalign}
		
		\begin{theorem}
			Optimization problem \eqref{prob:opt2} is equivalent \eqref{prob:opt1}. In other words, decreasing the worst case centrality index is equivalent to a network with the best robustness against an adversarial attack.
		\end{theorem}

		\subsection{Centrality rankings in special graph structures}\label{structure_centrality}
		
		\begin{proposition}
			In a network with a vertex-transitive coupling graph, e.g., complete graph and ring graph with uniform weight, agent centrality index $\eta_i = \bar{\rho}/n$, where $\bar\rho $ is the performance of the network with $\sigma_1 = \sigma_2 = \dots = \sigma_n = 1$. Similarly, in a network with an edge-transitive coupling graph, e.g. complete (bipartite) graph, and ring graph, link centrality index  $\nu_e = \dbar{\rho}/\mid\NHS \EE\NHS\mid$, where $\dbar\rho $ is the performance of the network with $\sigma_1 = \sigma_2 = \dots = \sigma_{\mid\EE\mid} = 1$. 
		\end{proposition}

		\begin{proposition}
			In a network with a tree graph with uniform weight $\bar{w}$, in the absence of time-delay, $\nu_e = 1/\bar{w}$ for all links. 
		\end{proposition}
		\begin{figure}[t]
			\centering
			\includegraphics[width=0.46\textwidth]{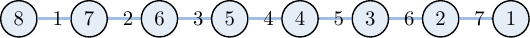}
			\caption{A path graph with labeled agents and links discussed in subsection \ref{structure_centrality}.}
			\label{fig_path}
		\end{figure}
		\begin{figure}[t]
			\centering
			\includegraphics[width=0.46\textwidth]{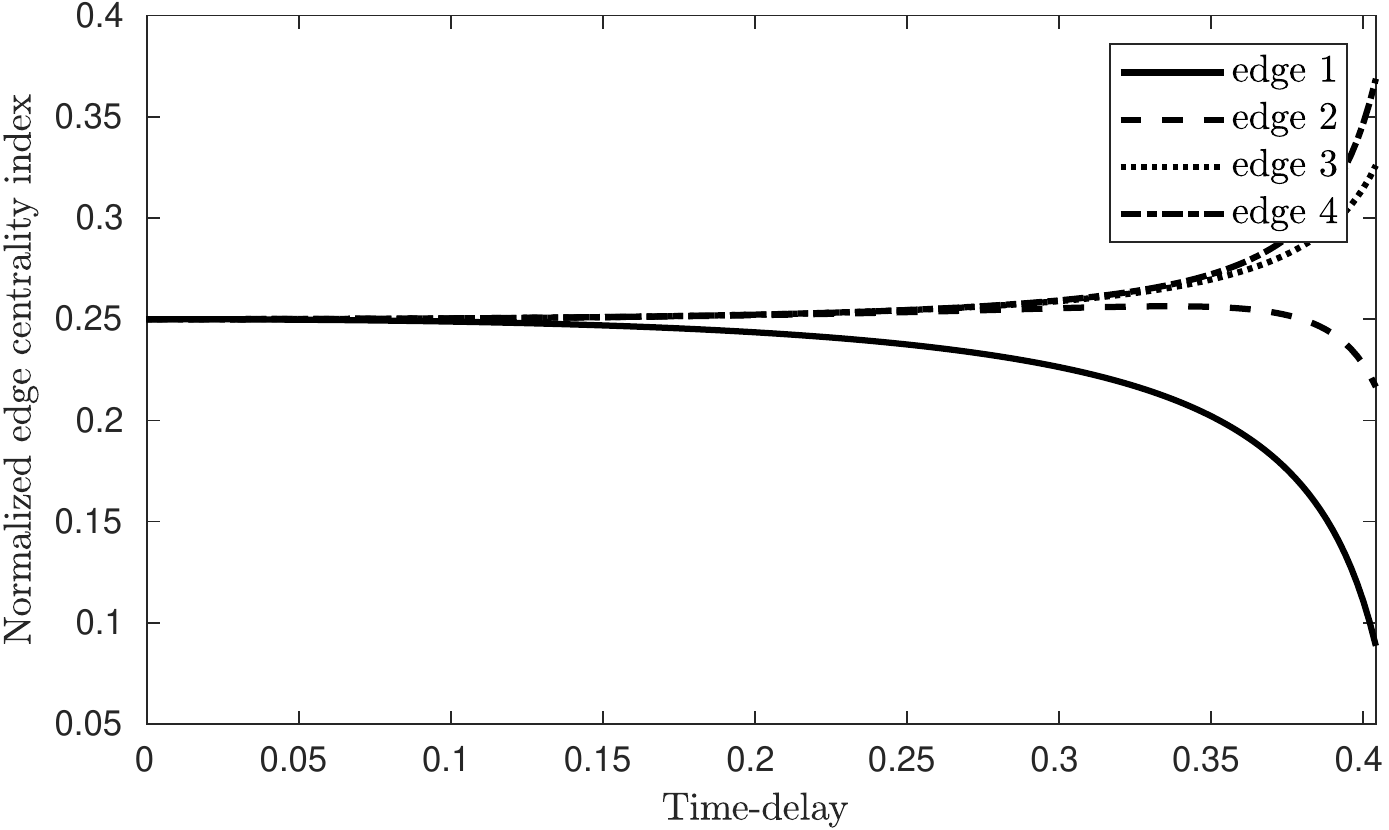}
			\caption{\footnotesize{Normalized link centrality in a path graph. Edge labels are provided in Fig \ref{fig_path}.}}     \label{Fig:cent_edge_path}
		\end{figure}
		
		\begin{figure}[t]
			\centering
			\includegraphics[width=0.46\textwidth]{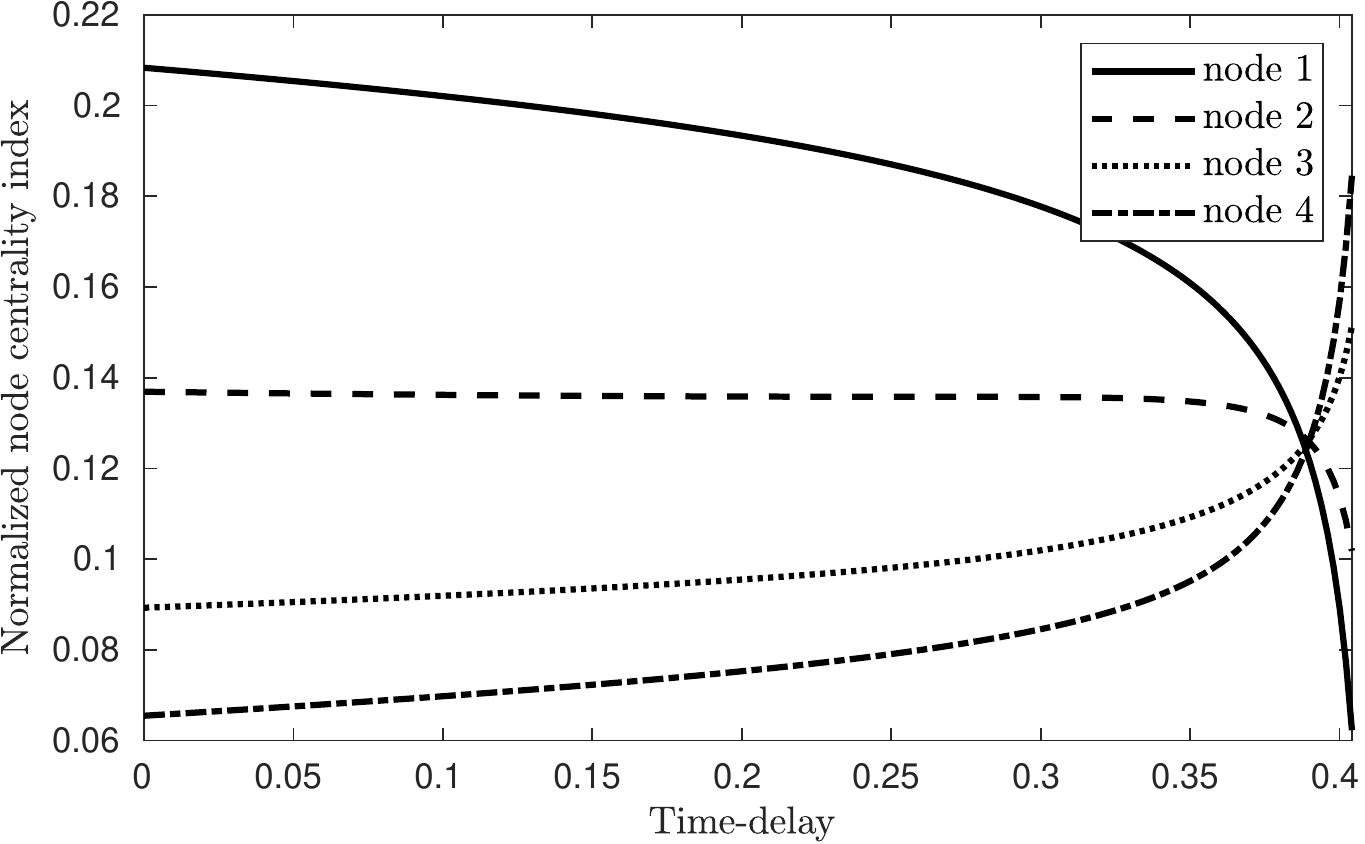}
			\caption{\footnotesize{Normalized agent centrality in a path graph with dynamics noise.}}     \label{Fig:cent_node_path}
		\end{figure}

		Figure \ref{Fig:cent_edge_path} depicts the centrality for a path graph with 8 agents in the presence of time-delay. We can see that initially ($\tau = 0$) all the links have equal centrality index, however, as time-delay increases, the central index of inner links increase with a higher rate than the outer links. In Figure \ref{Fig:cent_node_path}, centrality index of agents in a path graph with dynamics noise as a function time-delay is depicted. We can see that in the absence of time-delay, outer agents have higher centrality index. However, as time-delay increases, the inner agents gain higher centrality index.
		
		\begin{figure}[t]
			\centering
			\includegraphics[width=0.45\textwidth]{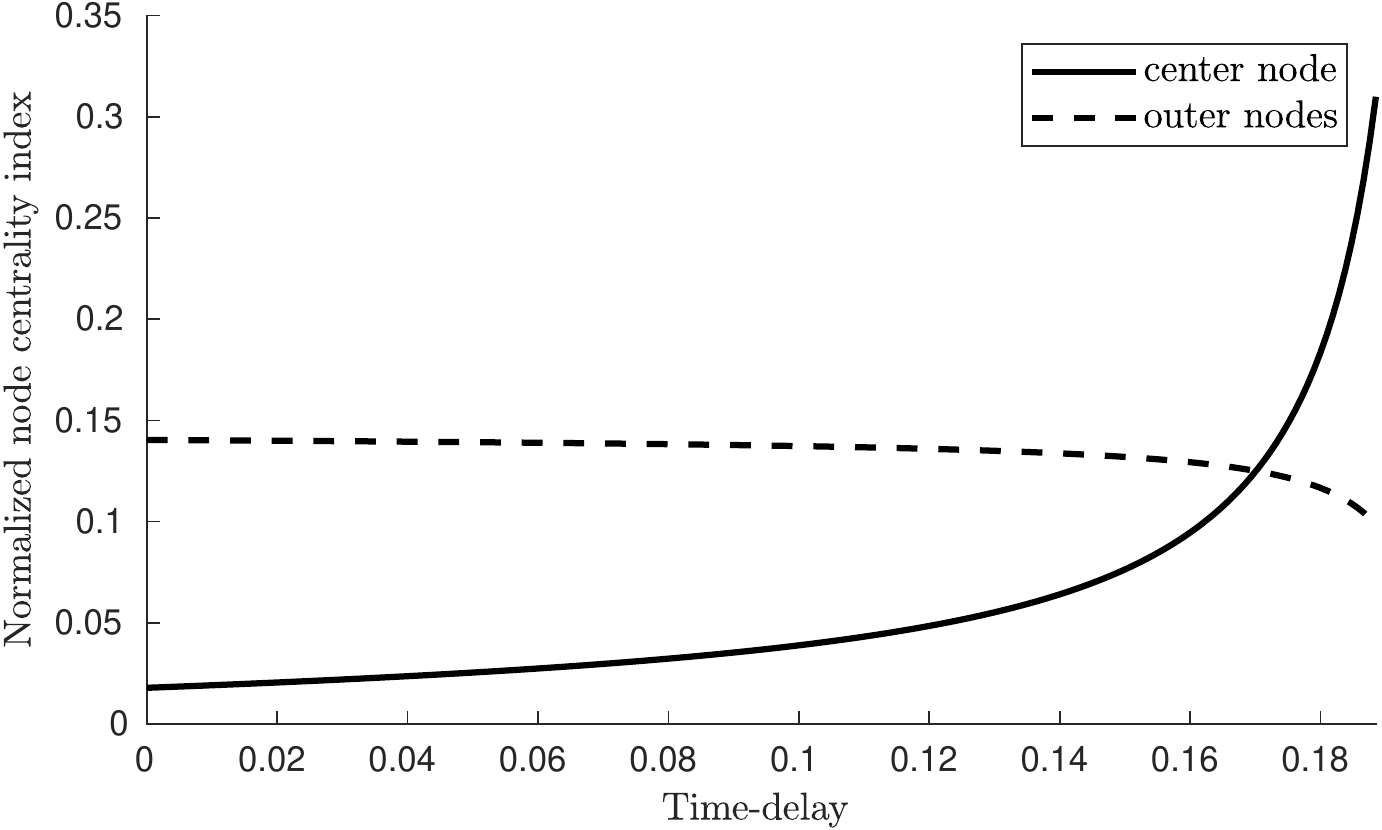}
			\caption{\footnotesize{Normalized agent centrality in a star graph with dynamics noise.}}     \label{Fig:cent_node_star}
		\end{figure}
		
	}

	\section{Conclusion}
	Interpretations of centrality and sensitivity measures, with respect to the $\mathcal H_2$-norm square, are proposed for networks with consensus dynamics subject to time-delays and structured additive noise inputs. In such networks, the centrality/sensitivity of each agent/communication link depends on the coupling graph of the network, time-delays and the structure of noise input.  
	We consider  several  uncertainty  structures that have real-world interpretations. It is shown that the centrality and sensitivity ranks of agents or links may vary substantially when comparing various noise structures and the time-delay. }

\bibliographystyle{plain}
{\bibliography{Ref/Bib.bib}}




\end{document}

%% file: dynamicNoise.tex
\tikzstyle{block} = [draw, fill=blue!10, rectangle, 
    minimum height=3em, minimum width=3em]
\tikzstyle{sum} = [draw, fill=blue!10, circle, node distance=1cm]
\tikzstyle{input} = [coordinate]
\tikzstyle{output} = [coordinate]
\tikzstyle{pinstyle} = [pin edge={to-,thin,black}]

\begin{tikzpicture}[auto, node distance=2cm,>=latex']
    \node [input, name=input] {};
    \node [sum, right of=input] (sum) {};
    \node [block, right of=sum] (integrator) {$\int$};
    \node [block, right of=integrator,            node distance=3cm] (system) {Delay};
    \draw [->] (integrator) -- node[name=x] {$x(t)$} (system);
    \node [output, right of=system] (output) {$x(t-\tau)$};
    \node [block, below of=x] (feedback) {$-L$};

    \draw [draw,->] (input) -- node {$\xi(t)$} (sum);
    \draw [->] (sum) -- node { } (integrator);
    \draw (6.5,0) -| (7.8,-1.7);
    \draw[->] (7.8,-1.7) -- (feedback)node[midway, below](TextNode) {$x(t-\tau)$};
    \draw [->] (feedback) -|  node[pos = 0.99] {$+$}
        node [near end] {} (sum);
\end{tikzpicture}

%% file: snsrNoise.tex
\tikzstyle{block} = [draw, fill=blue!10, rectangle, 
    minimum height=3em, minimum width=3em]
\tikzstyle{sum} = [draw, fill=blue!10, circle, node distance=2cm]
\tikzstyle{upperLeft} = [coordinate]
\tikzstyle{output} = [coordinate]
\tikzstyle{pinstyle} = [pin edge={to-,thin,black}]

\begin{tikzpicture}[auto, node distance=2cm,>=latex']
    \node [upperLeft, name=upperLeft] {};
    \node [upperLeft, name=upperRight] {};
    
    \node [block, right of=upperLeft] (integrator) {$\int$};
    \node [block, right of=integrator, node distance=3cm] (system) {Delay};

    \draw [->] (integrator) -- node[name=x] {$x(t)$} (system);
   
    \node [block, below of=x] (feedback) {$-L$};

    \node (upLeft) [upperLeft, right = 2cm of system] {};
    \node (sum) [sum, right  = 3.5cm of  feedback] {};
    \node (input) [upperLeft, right  = 0.5cm of  sum] {};

    \draw [draw,->] (input) -- node {$\xi(t)$} (sum);
    \draw (system) -- (upLeft)node[midway, above](TextNode) {$x(t-\tau)$};
    \draw[->] (upLeft) -| node[pos = 0.99] {$+$}(sum);
    \draw [->] (sum) -- (feedback);
    \draw [-] (feedback) -| 
        (1,0) ;
    \draw [->] (1,0) -- 
        (integrator);
    
\end{tikzpicture}

%% file: comNoise.tex
\tikzstyle{block} = [draw, fill=blue!10, rectangle, 
    minimum height=3em, minimum width=3em]
\tikzstyle{sum} = [draw, fill=blue!10, circle, node distance=2cm]
\tikzstyle{upperLeft} = [coordinate]
\tikzstyle{output} = [coordinate]
\tikzstyle{pinstyle} = [pin edge={to-,thin,black}]

\begin{tikzpicture}[auto, node distance=2cm,>=latex']
    \node [upperLeft, name=upperLeft] {};
    \node [upperLeft, name=upperRight] {};
    
    \node [block, right of=upperLeft] (integrator) {$\int$};
    \node [block, right of=integrator, node distance=3cm] (system) {Delay};

    \draw [->] (integrator) -- node[name=x] {$x(t)$} (system);
   
    \node [block, below of=integrator] (feedback) {$-EW$};

    \node (upLeft) [upperLeft, right = 2cm of system] {};
    
    \node (sum) [sum, right  = 1cm of  feedback] {};
    \node [block,right of=feedback, node distance=4cm] (feed1) {$E^{\rm T}$};
    \node (input) [upperLeft, below  = 0.5cm of  sum] {};

    \draw [draw,->] (input) -- node {$\xi(t)$} (sum);
    
    \draw (system) -- (upLeft)node[midway, above](TextNode) {$x(t-\tau)$};
    
    \draw[->] (upLeft) |- (feed1);
    \draw[->] (feed1) -- node[pos = 0.99] {$+$}(sum);
    \draw [->] (sum) -- (feedback);
    \draw [-] (feedback) -| 
        (1,0) ;
    \draw [->] (1,0) -- 
        (integrator);
    
\end{tikzpicture}

%% file: measNoise.tex
\tikzstyle{block} = [draw, fill=blue!10, rectangle, 
    minimum height=3em, minimum width=3em]
\tikzstyle{sum} = [draw, fill=blue!10, circle, node distance=2cm]
\tikzstyle{upperLeft} = [coordinate]
\tikzstyle{output} = [coordinate]
\tikzstyle{pinstyle} = [pin edge={to-,thin,black}]

\begin{tikzpicture}[auto, node distance=2cm,>=latex']
    \node [upperLeft, name=upperLeft] {};
    \node [upperLeft, name=upperRight] {};
    
    \node [block, right of=upperLeft] (integrator) {$\int$};
    \node [block, right of=integrator, node distance=3cm] (system) {Delay};

    \draw [->] (integrator) -- node[name=x] {$x(t)$} (system);
   
    \node [block, below of=integrator] (feedback) {$-E$};

    \node (upLeft) [upperLeft, right = 2cm of system] {};
    
    \node (sum) [sum, right  = 1cm of  feedback] {};
    \node [block,right of=feedback, node distance=4cm] (feed1) {$W E^{\rm T}$};
    \node (input) [upperLeft, below  = 0.5cm of  sum] {};

    \draw [draw,->] (input) -- node {$\xi(t)$} (sum);
    
    \draw (system) -- (upLeft)node[midway, above](TextNode) {$x(t-\tau)$};
    
    \draw[->] (upLeft) |- (feed1);
    \draw[->] (feed1) -- node[pos = 0.99] {$+$}(sum);
    \draw [->] (sum) -- (feedback);
    \draw [-] (feedback) -| 
        (1,0) ;
    \draw [->] (1,0) -- 
        (integrator);
    
\end{tikzpicture}